\documentclass[preprint,nofootinbib,floatfix,a4paper,prd,superscriptaddress]{revtex4-1}   	
\usepackage{geometry}                		
\usepackage[colorlinks,citecolor=blue]{hyperref}
\usepackage{amsmath}
\usepackage{mathrsfs}
\usepackage{bbm}
\usepackage{amsfonts}
\usepackage{pifont}
\usepackage{amssymb}
\usepackage{latexsym}
\usepackage{graphicx}
\usepackage[english]{babel}
\usepackage{multirow}
\usepackage{float}
\usepackage{url}
\usepackage{slashed}
\usepackage{xcolor} 
\usepackage{orcidlink}
\usepackage{appendix}

\newcommand{\be}{\begin{equation}}
\newcommand{\ee}{\end{equation}}
\newcommand{\ba}{\begin{array}}
\newcommand{\ea}{\end{array}}
\newcommand{\bea}{\begin{eqnarray}}
\newcommand{\eea}{\end{eqnarray}}

\usepackage[utf8]{inputenc}
\usepackage[T1]{fontenc}

\def  \bcen   {\begin{center}}
\def  \ecen   {\end{center}}
\def  \beq    {\begin{equation}}
\def  \eeq    {\end{equation}}
\def  \nn     {\nonumber }

\def\nn{\nonumber}

\def\to {\rightarrow}

\newcommand{\TSUa}{\affiliation{\small Tsung-Dao Lee Institute \& School of Physics and Astronomy, Shanghai Jiao Tong University, Shanghai 200240, China }}

\newcommand{\PU}{\affiliation{\small Faculty of Fundamental Sciences, PHENIKAA University, Yen Nghia, Ha Dong, Hanoi 12116, Vietnam}}

\newcommand{\AS}{\affiliation{\small Institute of Physics, Academia Sinica, Nangang, Taipei 11529, Taiwan}}

\begin{document}

\title{Self-interacting Vectorial Dark Matter \\in a SM-like Dark Sector}

\author{Van Que Tran \orcidlink{0000-0003-4643-4050}}
\email{vqtran@sjtu.edu.cn} \TSUa \PU

\author{Thong T. Q. Nguyen \orcidlink{0000-0002-8460-0219}}
\email{ntqthonghep@gate.sinica.edu.tw} \AS

\author{Tzu-Chiang Yuan \orcidlink{0000-0001-8546-5031} \, }
\email{tcyuan@phys.sinica.edu.tw} \AS

\begin{abstract}

A $SU(2)_D \times U(1)_D$ gauge-Higgs sector, an exact dark copy of the Standard Model (SM) one, is proposed. It is demonstrated that the
dark gauge bosons $\mathcal W^{(p,m)}$, in analogous to the SM $W^\pm$, 
can fulfill the role as a self-interacting vector dark matter candidate, solving the core versus cusp and missing satellites problems faced by the conventional paradigm of collisionless weakly interacting massive particle. Constraints from collider, astroparticle and cosmology on such a self-interacting vector dark matter candidate are scrutinized. Implications for the future searches of 
$\mathcal W^{(p,m)}$ in direct detection experiments are discussed.

\end{abstract}

\maketitle


\section{Introduction}
\label{Intro}

The very conception of standard model of cosmology is the 
$\Lambda$CDM paradigm in which the total mass-energy content of the Universe consists approximately 26\% of dark matter (DM), 5\% of baryonic matter and the rest is consistently described by a minuscule cosmological constant $\Lambda$. The DM particles in $\Lambda$CDM cosmology is cold, collisionless and interact purely gravitationally with each other and Standard Model (SM) particles. On large scales greater than $\mathcal{O}({\rm Mpc})$, the structure of the Universe is well described by the $\Lambda$CDM cosmology, as reflected in the predictions of DM abundance, acoustic peaks in the microwave background radiation, matter power spectrum, structure formation, {\it etc}. However, since 1990s various puzzles had been posed on smaller scales for the cold dark matter (CDM) predictions. Notably they are the core versus cusp, missing satellites and too-big-to-fail problems. These are phrased as small scale ``crisis'' in the literature.

While SM does not have any DM candidate, many models beyond the SM (BSM) 
can provide such a weakly-interacting massive particle (WIMP) to be the CDM. Most popular models include the neutralino in the minimal supersymmetric standard model (MSSM) motivated by the gauge hierarchy problem as well as the extremely light axion particle associated with the solution of the strong CP problem in QCD. Nevertheless all these DM candidates in the WIMP scenario suffer from the small scale ``crisis'' mentioned in previous paragraph. 

Over the years, many efforts were made to solve the 
small scale ``crisis'' within the $\Lambda$CDM paradigm. There were (and still are) high hopes that by including the baryon feedback processes such as gas cooling, star formation, supernovae, and active galactic nuclei in the $N$-body simulations, the ``crisis'' can be resolved. To the best of our knowledge, as of today, this issue is not completely settled.

Alternatively one may want a paradigm shift by introducing self-interacting dark matter (SIDM) first proposed by Spergel and Steinhardt~\cite{Spergel:1999mh} some times ago as a solution to the core-cusp and missing satellites problems. The idea is simple -- 
SIDM is collisional and thus conducts heat. Self-interactions can cause energy transfer from the hotter outer region to the colder central region, thereby forming a core. As the core of a galaxy heats up it expands, therefore decreasing the central density.  

The simplest SIDM particle candidate is to augment the SM by adding a scalar gauge singlet $\chi$ with a discrete symmetry $\chi \to -\chi$ that allows only a quartic $\chi^4$ self-coupling and a quartic $\chi^2 \Phi_H^\dagger \Phi_H$ coupling where $\Phi_H$ is the SM Higgs doublet as discussed in~\cite{Silveira:1985rk,Holz:2001cb,Cheung:2012xb}. At low energy, the cross section $\sigma_{\chi\chi \to \chi\chi}$ is a constant (velocity independent), dominated by the $\chi^4$ coupling while the $t$ channel exchange diagram due to the 125 GeV SM Higgs is negligible.  To solve the small scale ``crisis'' for dwarf galaxies, the SIDM cross section must be in the following range~\cite{Spergel:1999mh},
\be
\label{SpergelStienhardtBound}
0.5 \, {\rm cm ^2/g} \lesssim \sigma/m_\chi \lesssim 6 \, \mathrm{cm^2/g} \; ,
\ee
which implies~\cite{Holz:2001cb}
\be
m_\chi \sim \left( 0.03 - 0.08 \right) \lambda_\chi^{2/3} \, \mathrm{GeV} \; ,
\ee
where $\lambda_\chi$ is the $\chi^4$ coupling. Thus $\chi$ is a sub-GeV SIDM if its self-coupling is of order one.

In recent years, there is a resurgence of interest in SIDM due to the series of works in~\cite{Tulin:2013teo,Kaplinghat:2015aga,Tulin:2017ara,Colquhoun:2020adl}. Astrophysical data accumulated over the past several decades for dwarf galaxies, low surface brightness (LSB) spiral galaxies and galaxy clusters indicating DM self-interaction cross sections can have significant velocity dispersion. The range in~(\ref{SpergelStienhardtBound}) has been extended to~\cite{Tulin:2017ara} 
\be
\label{SIDMxsectionBound}
0.1 \, {\rm cm ^2/g} \lesssim \sigma/m_\chi \lesssim 10 \, {\rm cm^2/g} \; ,
\ee
depending on the average velocity of the DM $\chi$ in the observed astrophysical objects. We will return to this issue in later section. For details on the small scale ``crisis'', baryon feedback and SIDM, the readers are redirected to a recent review by Tulin and Yu~\cite{Tulin:2017ara}.

Many BSM models can give rise to non-negligible $t$ channel exchange contribution to SIDM cross section with mediator mass much lighter than the DM mass, like the dark $U(1)$ force interacts with a fermionic DM in~\cite{Aboubrahim:2020lnr}. Moreover the SIDM cross sections in this class of BSM models have nontrivial velocity dependent of the DM and have sizes in the desirable range given in (\ref{SIDMxsectionBound}).


\begin{table}[t!]
\begin{tabular}{|c|c|c||c|c|c|c|}
\hline
Scalar Fields  & $SU(2)_L$ & $U(1)_Y$ &  $SU(2)_D$ & $U(1)_D$ \\
\hline\hline
$\Phi_H$ & $2$ & $\frac{1}{2}$ & $1$ & $0$ \\
$\Phi_D$ & $1$ & $0$ & $2$ & $\frac{1}{2}$ \\
\hline
\end{tabular}
\caption{Scalar field contents in the model. Our convention is 
$Q = T^3 + Y$ where $Q$ is the electric charge in unit of $e > 0$, $T^3$ is the third component of the generators of $SU(2)_L$ 
and  $Y$ is the hypercharge. 
}
\label{tab:quantumnos}
\end{table}

In this paper, motivated by our earlier works on gauged two-Higgs-doublet model (G2HDM)~\cite{Huang:2015wts} and its minimal version~\cite{Ramos:2021txu,Ramos:2021omo}, we consider a $SU(2)_D \times U(1)_D$ gauge-Higgs sector which is exactly a dark copy of the SM one $SU(2)_L \times U(1)_Y$. 
The dark gauge bosons $\mathcal W^{(p,m)}$ of $SU(2)_D$, in parallel with the SM $W^\pm$, 
can be a vectorial dark matter candidate. The main motivation or advantage is that,  due to the non-abelian cubic and quartic couplings, $\mathcal W^{(p,m)}$ is naturally self-interacting. Furthermore, since $\mathcal W^{(p,m)}$ is complex, one can have either symmetric (with both $\mathcal W^p$ and $\mathcal W^m$) or asymmetric (with either $\mathcal W^p$ or $\mathcal W^m$) DM scenario~\cite{Graesser:2011wi,Iminniyaz:2011yp,Lin:2011gj,Zurek:2013wia}. 
We will focus on the symmetric case in what follows. The fermion content is the same as in SM, while the scalar fields used in the model are shown in Table~\ref{tab:quantumnos}.
The communication between the SM and dark sectors are via three portals:
\begin{itemize}
    \item Kinetic mixing of the two abelian groups, 
the hypercharge $U(1)_Y$ in SM and $U(1)_D$ in the dark sector, 
\item Higgs mixing among the SM Higgs doublet $\Phi_H$ and its dark partner $\Phi_D$, and
\item Mass mixings among the three neutral gauge bosons.
\end{itemize}
A Stueckelberg mass is introduced for 
$U(1)_D$ such that all gauge bosons are massive except the SM photon.
We will be interested SIDM in the sub-GeV to TeV mass range with light mediators in order to resolve some of the well-known issues like core-cusp and too-big-to-fail problems~\cite{Tulin:2017ara} that one faces 
in the more conventional WIMP CDM scenario. 
Indeed, the model consists of one dark photon $\gamma^\prime$, one dark $Z^\prime$ and one dark Higgs, all of which can play the role of a mediator $X$ for SIDM. These extra particles 
can have important impacts in collider physics as well. 

Computation of the SIDM cross section is highly nontrivial as pointed out in~\cite{Colquhoun:2020adl}. Methods used for the computation depend on two dimensionless parameters $\kappa$ and $\beta$ constructed from the following 4 quantities: coupling strength ($\alpha_X$) between the SIDM ($\mathcal W$) and its mediator ($X$), their masses ($m_{\cal W}$ and $m_{X}$) and the velocity $v$ of the DM, 
\beq 
\label{eq:kappaandbeta}
\kappa = \frac{m_{\cal W} v}{m_{X}} \; , \quad\quad  \beta=\frac{2\alpha_X m_{X}}{m_{\cal W}v^2 } \; .
\eeq
The na\"{i}ve Born approximation ($2\beta\kappa^2 \ll 1$) based on perturbative expansion in $\alpha_X$ would break down when the mediator mass is much lighter than the DM mass 
resulting a long range force. 
Non-perturbative effects must be taken into account in the 
semi-classical ($\kappa \geq 1$) and quantum regimes ($\kappa \ll 1$),
as suggested in~\cite{Colquhoun:2020adl} which we will follow closely in this work.

The content of this paper is unfolded as follows. 
In section~\ref{The Model}, we present the Lagrangian of the gauge-Higgs sector in the model and discuss its mass spectra after spontaneously symmetry breaking. Phenomenological constraints for the vectorial DM $\mathcal W^{(p,m)}$ from collider physics, electroweak precision tests, dark photon and dark $Z^\prime$ physics, relic density and direct detection are considered in section~\ref{pheno}. Implementation of $\mathcal W^{(p,m)}$ as SIDM is discussed in section~\ref{sec:SIDM}, followed by our numerical analysis in section~\ref{numbers}. We conclude in section~\ref{conclusions}. 
Details of the relevant couplings in the neutral current interaction Lagrangian for the fermions and gauge bosons 
are given in appendix~\ref{app:NCI}.
Results of an alternative mass hierarchy of the dark photon and dark $Z^\prime$ is given in appendix~\ref{app:result_case2}.



\section{The $SU(2)_L \times U(1)_Y \times SU(2)_D \times U(1)_D$ Model}
\label{The Model}

As alluded to in previous section, the main advantage of having a non-abelian gauge-Higgs dark sector of 
$SU(2)_D \times U(1)_D$ is that the extra gauge bosons $\mathcal W^{(p,m)}$ 
can be naturally a self-interacting dark matter candidate without the need of introducing other matter field.
Thus the Lagrangian of the gauge sector is given by 
\begin{align}\label{L-gauge}
{\mathcal L_G}={}& -\frac{1}{4} F^a_{\mu\nu} F^{a{}\mu\nu} - \frac{1}{4} B_{\mu\nu} B^{\mu\nu}
-\frac{1}{4} \mathcal F^a_{\mu\nu} \mathcal F^a_{\mu\nu} - \frac{1}{4} \mathcal X_{\mu\nu} \mathcal X^{\mu\nu} \nn \\
&-\frac{1}{2} \epsilon \mathcal X_{\mu\nu}B^{\mu\nu} + \frac{1}{2} \left( \partial_\mu \sigma + M_{\mathcal X} {\mathcal X}_\mu \right)^2 + \mathrm{gauge \; fixings} \; ,
\end{align} 
where $F^a_{\mu\nu}$ and $\mathcal F^a_{\mu\nu}$ $(a=1,2,3)$ are the $SU(2)_L$ and $SU(2)_D$ field strength tensors respectively,
$M_{\mathcal X}$ is the Stueckelberg mass~\cite{Kors:2004dx,Kors:2005uz,Cheung:2007ut}
for the $U(1)_D$ gauge field ${\mathcal X}_\mu$
associated with its field strength $\mathcal X_{\mu\nu}$, and
$\epsilon$ is the kinetic mixing parameter~\cite{Holdom:1985ag,Feldman:2007wj} between the SM hypercharge field strength $B_{\mu\nu}$ and $\mathcal X_{\mu\nu}$.
$\sigma$ is the auxiliary scalar field introduced to implement the Stueckelberg mechanism to provide a mass for the $U(1)_D$ gauge boson in a gauge invariant fashion.
The SM fermions are the same and they are inert under the dark group $SU(2)_D \times U(1)_D$. Dark fermions are also possible but we do not introduce them to keep thing as minimal as possible.

On the other hand, the Lagrangian for the scalar sector is given by
\begin{align}\label{L-Higgs}
{\mathcal L}_S = {}& \vert D_\mu \Phi_H \vert^2 +  \vert D_\mu \Phi_D \vert^2 - V\left( \Phi_H , \Phi_D \right) \; ,
\end{align}
where the covariant derivatives of $D_\mu \Phi$ and $D_\mu \Phi_D$ are given by the familiar expressions~\footnote{Here for  $\Phi_H$
we have $T^\pm = T^1 \pm i T^2 = \left( \sigma^1 \pm i \sigma^2 \right) /2$ and 
$T^3 = \sigma^3/2$ where $\sigma^i (i=1,2,3)$ are the Pauli matrices.
Similarly for $\Phi_D$ we have $\mathcal T^\pm = \mathcal T^1 \pm i \mathcal T^2 = \left( \sigma^1 \pm i \sigma^2 \right) /2$ and $\mathcal T^3 = \sigma^3/2$.
$W^\pm_\mu = \left( W^1_\mu \mp i W^2_\mu \right)/\sqrt 2$ and $W^3_\mu$ are the gauge fields of $SU(2)_L$,
$\mathcal W^{(p,m)}_\mu = \left( \mathcal W^1_\mu \mp i \mathcal W^2_\mu \right)/\sqrt 2$ and $\mathcal W^3_\mu$ are the gauge fields of $SU(2)_D$, 
and finally $B_\mu$ and $\mathcal X_\mu$ are the gauge fields of $U(1)_Y$ and $U(1)_D$ respectively.
}
\begin{align}
D_\mu \Phi_H = & \left[ \partial_\mu - i \frac{g}{2} \left( W^+_\mu T^+ + W^-_\mu T^- \right) - i g W_\mu^3 T^3 - i g^\prime \frac{1}{2} B_\mu \right] \Phi_H \; ,\nn \\
D_\mu \Phi_D = & \left[ \partial_\mu - i \frac{g_D}{2} \left( {\mathcal W}^p_\mu {\mathcal T}^p + {\mathcal W}^m_\mu {\mathcal T}^m \right) - i g_D {\mathcal W}^3_\mu \mathcal T^3 - i g_D^\prime \frac{1}{2} 
{\mathcal X}_\mu \right] \Phi_D \; ;
\end{align}  
and
$V(\Phi_H,\Phi_D)$ is the most general Higgs potential invariant under both $SU(2)_L\times U(1)_Y$ and  $SU(2)_D \times  U(1)_D$  
\begin{align}\label{eq:V}
V \left( \Phi_H , \Phi_D \right) = {}& - \mu^2    \Phi_H^\dagger  \Phi_H  
+  \lambda \left( \Phi_H^\dagger  \Phi_H \right)^2  
- \mu^2_D   \Phi_D^\dagger  \Phi_D  
+  \lambda_D \left( \Phi_D^\dagger  \Phi_D \right)^2  \nn \\
&+ \lambda^{\prime}\left(  \Phi_H^\dagger  \Phi_H \right) \left( \Phi_D^\dagger  \Phi_D \right) \; .
\end{align}

Neutral current interaction Lagrangian for the fermions and gauge bosons can be found in appendix~\ref{app:NCI}.

\subsection{Spontaneous Symmetry Breaking}
\label{HiggsPotential}

To study the spontaneous symmetry breaking (SSB) in the model, we decompose the fields in the two scalar doublets according to 
\begin{eqnarray}
\label{eq:scalarfields1}
\Phi_H & = & 
\begin{pmatrix}
G^+ \\ \frac{1}{\sqrt 2} \left( v_H + h + i G^0 \right)
\end{pmatrix}
\;, \;\; 
\Phi_D = 
\begin{pmatrix}
G^p \\ \frac{1}{\sqrt 2} \left( v_D + h_D + i G^0_D \right)
\end{pmatrix}
\; .
\end{eqnarray}
Here $v_H = 246$ GeV is the SM vacuum expectation value (VEV), and $v_D$ is the unknown hidden VEV,
\begin{eqnarray}
\label{eq:scalarfields}
\langle \Phi_H \rangle =
\begin{pmatrix}
0 \\ \frac{v_H}{\sqrt 2}
\end{pmatrix}
, \;
\langle \Phi_D \rangle = 
\begin{pmatrix}
0 \\ \frac{v_D }{\sqrt 2}
\end{pmatrix}
. \;
\end{eqnarray}
The potential for the VEVs is then
\beq
V (v_H, v_D) =  \frac{1}{4}\left(  -2 \mu^2 v_H^2  + \lambda v_H^4  - 2 \mu_D^2 v_D^2 + \lambda_D v_D^4  + \lambda^{\prime}  v_H^2 v_D^2 \right) \; .
\eeq
To determine the true vacuum, we impose the tadpole conditions $\partial V/\partial v_H = 0$ and $\partial V/\partial v_D = 0$, which lead to
\begin{align}
\label{TP1}
& v_H \left( - 2 \mu^2 + 2 \lambda v_H^2  + \lambda^{\prime} v_D^2 \right)  = 0 \; , \\
\label{TP2}
& v_D \left( - 2 \mu_D^2 + 2 \lambda_D v_D^2 + \lambda^\prime v_H^2 \right) = 0 \; .
\end{align}
Non-trivial solutions for the VEVs are
\begin{align}
v_H^2 =& \frac{4 \lambda_D \mu^2 - 2 \lambda^\prime \mu_D^2 }{4 \lambda \lambda_D - \lambda^{\prime 2}} \; , \\
v_D^2 =& \frac{4 \lambda \mu_D^2 - 2 \lambda^\prime \mu^2 }{4 \lambda \lambda_D - \lambda^{\prime 2}} \; .
\end{align}

Besides requiring the vacuum to be non-tachyonic 
$v_H^2 > 0$ and $v_D^2 > 0$, to trust our perturbative calculations, we must impose
\beq
\vert \lambda \vert, \; \vert \lambda_D \vert  \; \textrm{and}  \; \vert \lambda^\prime \vert \leq 8 \pi \; .
\eeq
Tree level perturbative unitarity in the scalar sector implies further that~\cite{Arhrib:2012ia,Arhrib:2013ela}
\beq
\vert \lambda \vert \; \textrm{and} \; \vert \lambda_D \vert \leq \frac{4 \pi}{3} \; .
\eeq
Moreover, requiring the potential bounded from below implies~\cite{Branco:2011iw}
\beq
\lambda > 0, \; \lambda_D > 0 \; \textrm{and} \;2 \sqrt{ \lambda \lambda_D} + \lambda^\prime > 0 \; .
\eeq

\subsection{Mass Spectra}
\label{massspectra}


\subsubsection{Scalar Bosons}
\label{scalarmasses}

We will work in the 't Hooft-Landau gauge in which the Goldstone bosons from both SM and dark sectors are massless (in view of the above tadpole conditions (\ref{TP1}) and (\ref{TP2}))
\bea
m_{G^\pm}^{ 2} & = & m_{G^0}^{ 2} = -\mu^2 +  \lambda v_H^2 + \frac{1}{2} \lambda^\prime v_D^2 = 0 \; , \\
m_{G^{(p,m)}}^{ 2} & =& m_{G^0_D}^{ 2} = -\mu_D^2 +  \lambda_D v_D^2 + \frac{1}{2} \lambda^\prime v_H^2 = 0 \; .
\eea
The remaining two components $h$ and $h_D$  from the two doublets $\Phi_H$ and $\Phi_D$ mix according to
\beq
{\mathcal M}^2_S = 
\begin{pmatrix}
-\mu^2 + 3 \lambda v_H^2 + \frac{1}{2} \lambda^\prime v_D^2 & \lambda^\prime v_H v_D \\
\lambda^\prime v_H v_D & -\mu_D^2 + 3 \lambda_D v_D^2 + \frac{1}{2} \lambda^\prime v_H^2
\end{pmatrix}
=
\begin{pmatrix}
2 \lambda v_H^2 & \lambda^\prime v_H v_D \\
\lambda^\prime v_H v_D & 2 \lambda_D v_D^2
\end{pmatrix} 
\; .
\eeq
In the last equality we have used again the tadpole  conditions to eliminate $\mu^2$ and $\mu_D^2$.
The physical states $h_1$ and $h_2$ are related to $h$ and $h_D$ by
\begin{align}
h = &  h_1 \cos \alpha  - h_2 \sin \alpha \;, \\
h_D = &  h_1 \sin \alpha  + h_2  \cos \alpha  \; , 
\end{align}
with the mixing angle
\beq\label{mixinganglealpha}
\sin 2 \alpha = \frac{\lambda^\prime v_H v_D}{\left[ \left( \lambda v_H^2 - \lambda_D v_D^2 \right)^2 + \lambda^{\prime 2} v_H^2 v_D^2 \right]^{1/2}} \; ,
\eeq
and their masses are given by
\beq\label{scalarmassesh1h2}
m_{h_1,h_2}^2 = \left( \lambda v_H^2 + \lambda_D v_D^2\right)  \mp \left[ \left( \lambda v_H^2 - \lambda_D v_D^2 \right)^2 + \lambda^{\prime 2} v_H^2 v_D^2 \right]^{1/2} \; .
\eeq
We will identify the lighter Higgs $h_1$ with mass $m_{h_1} = 125.25 \pm 0.17$ GeV~\cite{PDG2022} corresponds to the scalar Higgs boson observed at LHC.

One can invert the above formulas (\ref{mixinganglealpha}) and (\ref{scalarmassesh1h2}) to trade the fundamental parameters 
$\lambda$, $\lambda_D$ and $\lambda^\prime$ with the two physical masses 
$m_1^2$, $m_2^2$ and the mixing angle $\alpha$, namely
\begin{align}
\label{lambda}
\lambda = & \frac{1}{2 v_H^2} \left( m_{h_1}^2 \cos^2 \alpha + m_{h_2}^2 \sin^2\alpha \right)  \; , \\
\label{lambdaD}
\lambda_D = &  \frac{1}{2 v_D^2} \left( m_{h_1}^2 \sin^2 \alpha + m_{h_2}^2 \cos^2\alpha \right)  \; , \\
\label{lambdaprime}
\lambda^\prime = & \frac{1}{2 v_H v_D} \left[ \left( m_{h_2}^2 - m_{h_1}^2 \right) \sin 2 \alpha \right] \; .
\end{align}
These equations are useful for numerical studies.

\subsubsection{Gauge Bosons}
\label{gaugebosonmasses}

The mass spectrum of gauge bosons from spontaneous symmetry breaking can be obtained by looking at the terms in 
$\vert D_\mu \Phi_H \vert^2$  and  $\vert D_\mu \Phi_D \vert^2$ which are second order in the VEVs ($v_H,v_D$)  
as well as second order in the gauge couplings ($g,g^\prime,g_D,g_D^\prime$).

There is only one charged vector boson, the SM $W^\pm$, with mass given by the standard formula
\beq\label{massSMW}
m_{W^\pm} = \frac{1}{2} g  v_H   \; .
\eeq
Experimentally we have $m_W = 80.377 \pm 0.012 \; {\rm GeV}$~\cite{PDG2022}.

The complex dark gauge boson ${\mathcal W}^p$ (${\mathcal W}^m = ({\mathcal W}^p)^*$) carries one unit of $+(-)$ dark charge but no electric charge. 
In analogous to (\ref{massSMW}), their masses are  simply
\beq\label{massDarkW}
m_{\mathcal W^{(p,m)}} = \frac{1}{2} g_D  v_D   \; .
\eeq

All the other 4 gauge bosons $W^3, B,{\mathcal W^3}$ and $\mathcal X$ are electrically neutral and in general mixed together.
In the basis of
\beq\label{VN}
V_{N} = 
\begin{pmatrix}
 B \\ W^3 \\ \mathcal X \\ {\mathcal W}^3 
 \end{pmatrix} \; ,
 \eeq
one finds
\begin{eqnarray}
\label{MassV1}
{\mathcal M}_N^2 
&=& 
\begin{pmatrix}
\frac{1}{4} g^{\prime\, 2} v_H^2 & - \frac{1}{4} g g^\prime v_H^2 & 0 & 0 \\ 
- \frac{1}{4} g g^{\prime} v_H^2 &  \frac{1}{4} g^2 v_H^2 & 0 & 0 \\
0 & 0 &  \frac{1}{4} g_D^{\prime 2} v_{D}^2  + M_{\mathcal X}^2  &  - \frac{1}{4} g_D g_D^\prime v_{D}^2   \\
0 & 0 &  - \frac{1}{4} g_D g_D^{\prime} v_D^2 &  \frac{1}{4} g_D^2  v_D^2 
\end{pmatrix} \; .
\end{eqnarray}
Note that we have included a Stueckelberg mass $M_{\mathcal X}$ for the $U(1)_D$, while for the 
$U(1)_Y$ we set its Stueckelberg mass $M_Y$ to be zero.

Before diagonalize the above mass matrix ${\mathcal M}_N^2$, we need to first diagonalize the kinetic terms for $B_{\mu\nu}$ and $\mathcal X_{\mu\nu}$ 
due to their kinetic mixing. This can be accomplished by performing a $GL(4)$ transformation $K$ given by~\cite{Feldman:2007wj}
\beq
K = 
\begin{pmatrix}
1 & 0 & -S_\epsilon & 0 \\
0 & 1 & 0 & 0 \\
0 & 0 & C_\epsilon & 0 \\
0 & 0 & 0 & 1
\end{pmatrix}
\; , \;\;\; {\rm with} \; C_\epsilon = \frac{1}{\sqrt{1 - \epsilon^2}} \; {\rm and} \; S_\epsilon = \epsilon C_\epsilon \; .
\eeq
This induces a change in the mass matrix ${\mathcal M}_N^2 \to  {\mathcal M}_N^{\prime 2} = K^T \cdot {\mathcal M}_N^2 \cdot K $
and $V_N \to V^\prime_N = K^{-1} \cdot V_N$, \textit{i.e.}
\begin{align}\label{MassV2}
{\mathcal M}_N^{\prime 2} =  
\begin{pmatrix}
\frac{1}{4} g^{\prime\, 2} v_H^2 & - \frac{1}{4} g g^\prime v_H^2 & -\frac{1}{4} g^{\prime 2} v_H^2 S_\epsilon & 0 \\ 
- \frac{1}{4} g g^{\prime} v_H^2 &  \frac{1}{4} g^2 v_H^2 & \frac{1}{4} g g^{\prime} v_H^2 S_\epsilon & 0 \\
-\frac{1}{4} g^{\prime 2} v_H^2 S_\epsilon  & \frac{1}{4} g g^{\prime} v_H^2 S_\epsilon & \left(  \frac{1}{4} g_D^{\prime 2} v_{D}^2  + M_{\mathcal X}^2  \right) C^2_\epsilon
+ \frac{1}{4} g^{\prime 2} v_H^2 S^{2}_\epsilon &  - \frac{1}{4} g_D g_D^\prime v_{D}^2  C_\epsilon \\
0 & 0 &  - \frac{1}{4} g_D g_D^{\prime} v_D^2 C_\epsilon &  \frac{1}{4} g_D^2  v_D^2 
\end{pmatrix} \; ,
\end{align}
 and
 \beq\label{VNp}
V^\prime_{N} \equiv
\begin{pmatrix}
 B^\prime \\ W^{3 \prime} \\ \mathcal X^\prime \\ {\mathcal W}^{3 \prime}
 \end{pmatrix} =
 \begin{pmatrix}
 B + \epsilon \mathcal X\\ W^3 \\ \frac{1}{C_\epsilon} \mathcal X \\ {\mathcal W}^3 
 \end{pmatrix} \; .
 \eeq

Despite its unappealing look, (\ref{MassV2}) has zero determinant so it has at least one zero eigenvalue~\footnote{If $M_{\mathcal X}$ vanishes, (\ref{MassV2}) 
will have two zero eigenvalues. Besides the massless photon in the SM, one would have a massless dark photon as well.}.
Indeed, by using the following orthogonal transformation 
$V_N^\prime \to V_N^{\prime\prime} = O_W^T \cdot V^\prime$ where
\beq
O_W = 
\begin{pmatrix}
c_W & - s_W & 0 & 0 \\
s_W & c_W & 0 & 0 \\
0 & 0 & 1 & 0 \\
0 & 0 & 0 & 1 
\end{pmatrix} 
\;\;\; {\rm with} \; \; \left( c_W , s_W \right) = \frac{1}{\sqrt{g^2 + g^{\prime\, 2}}} \left( g , g^\prime \right) \; ,
\eeq
and hence~\footnote{Here $A_{\rm SM} = c_W B + s_W W^3$ and $Z_{\rm SM} = -s_W B + c_W W^3$ are the SM expressions.}
\beq
\label{VNpp}
V^{\prime\prime}_{N} \equiv
\begin{pmatrix}
 A \\ Z \\ \mathcal X^{\prime\prime} \\ {\mathcal W}^{3 \prime\prime} 
 \end{pmatrix}  =
 \begin{pmatrix}
 A_{\rm SM} + c_W \epsilon  \mathcal X   \\  Z_{\rm SM} -  s_W \epsilon  \mathcal X \\ \frac{1}{C_\epsilon} \mathcal X \\ {\mathcal W}^3  
 \end{pmatrix} \; ,
 \eeq
one can bring ${\mathcal M}_N^{\prime \,2}$ into the following form 
\begin{equation}
\label{OWTransformation}
O_W^T \cdot {\mathcal M}_N^{\prime \, 2}  \cdot O_W  = 
\begin{pmatrix}
0 & {\bf 0}^{\rm T} \\
{\bf 0} & \hat{\mathcal M}_N^2
\end{pmatrix}
\; ,
\end{equation}
where  $\hat{\mathcal M}_N^2$ is a reduced $3 \times 3$ matrix given by
\beq\label{MassV3}
 \hat{\mathcal M}_N^2 = \begin{pmatrix}
  \frac{1}{4} \left( g^2 + g^{\prime 2}\right) v_H^2 & \frac{1}{4} g^{\prime} \sqrt{g^2 + g^{\prime 2}} v_H^2 S_\epsilon & 0 \\ 
\frac{1}{4} g^{\prime} \sqrt{g^2 + g^{\prime 2}} v_H^2 S_\epsilon & \left(  \frac{1}{4} g_D^{\prime 2} v_{D}^2  + M_{\mathcal X}^2  \right) C^2_\epsilon
+ \frac{1}{4} g^{\prime 2} v_H^2 S_\epsilon^2 & - \frac{1}{4} g_D g_D^\prime v_{D}^2  C_\epsilon \\
0 &  - \frac{1}{4} g_D g_D^{\prime} v_D^2 C_\epsilon &  \frac{1}{4} g_D^2  v_D^2 
 \end{pmatrix} \; .
\eeq

One can see from (\ref{VNpp}) and (\ref{OWTransformation}) that the $A$ field has zero mass and can be identified as the physical photon in the model,
while the $Z$ field will be further mixed with $\mathcal X^{\prime\prime}$ and ${\mathcal W}^{3 \prime\prime}$ with the mass matrix defined in (\ref{MassV3}).
We thus perform a final orthogonal transformation $\mathcal O$ to diagonalize $\hat {\mathcal M}_N^2$ as follows
\beq
\begin{pmatrix}
Z \\ \mathcal X^{\prime\prime} \\ \mathcal W^{3 \prime \prime}
\end{pmatrix} 
\to 
\begin{pmatrix}
Z_1 \\ Z_2 \\ Z_3
\end{pmatrix} 
= \mathcal O^T \cdot 
\begin{pmatrix}
Z \\ \mathcal X^{\prime\prime} \\ \mathcal W^{3 \prime\prime}
\end{pmatrix} 
\; ,
\eeq
and 
\beq
\hat {\mathcal M}_N^2 \to \mathcal O^T \cdot \hat {\mathcal M}_N^2 \cdot \mathcal O 
= \begin{pmatrix}
m_1^2 & 0 & 0\\
0 & m_2^2 & 0 \\
0 & 0 & m_3^2
\end{pmatrix}
\; ,
\eeq
where $m_i^2$ is the mass-squared of the physical field $Z_i$ for $i=1,2,3$.
Note that ${\rm Det}\hat {\mathcal M}_N^2 = g_D^2 (g^2 + g_D^2) v_H^2 v_D^2 C_\epsilon^2  M_{\mathcal X}^2 / 16$. 
In the limit of $\epsilon \to 0$ and $M_{\mathcal X} \to 0$,  
it is clear from the structure of $\hat {\mathcal M}_N^2$ in (\ref{MassV3}) that
$Z_1 \to Z_{\rm SM}$ with $m_1^2 \to  m^2_{Z_{\rm SM}} = (g^2 + g^{\prime 2})^2 v_H^2/4$;
while $Z_2 \to \gamma_D = c_D \mathcal \chi^{\prime\prime} + s_D \mathcal W^{3 \prime\prime}$ and
$Z_3 \to \mathcal Z_D = -s_D \mathcal \chi^{\prime\prime} + c_D \mathcal W^{3 \prime\prime}$ 
with $m^2_2 \to m^2_{\gamma_D} = 0$ and $m^2_3 \to m^2_{\mathcal Z_D} = m^2_{\mathcal W^{(p,m)}} /c^2_D$, where 
$(c_D,s_D) = (g_D,g_D^\prime)/\sqrt{g_D^2 + g_D^{\prime 2}}$ 
in analogous to the SM Weinberg angle for the dark sector. 
In order to have a successful symmetry breaking pattern for general values of $\epsilon$ and $M_{\mathcal X}$,
it is necessary for $\hat {\mathcal M}_N^2$ to provide an eigenvalue with a square root within the experimental range of 
the physical $Z$ mass $m_Z= 91.1876 \pm 0.0021$ GeV~\cite{PDG2022}.
While it is possible to obtain analytic formulas for the rotation angles in $\mathcal O$ in terms of the fundamental parameters in the Lagrangian as was done 
in previous work~\cite{Huang:2019obt}, we will diagonalize $\hat {\mathcal M}_N^2$ numerically in this work.


\section{Phenomenological Constraints}
\label{pheno}

As mentioned in the Introduction section, the communication between the SM and dark sectors is controlled by the kinetic mixing parameter $\epsilon$, the Higgs mixing parameter $\lambda^\prime$, and the mass mixings among the three neutral gauge bosons.
By tuning these mixing effects sufficiently small, the SM physics remains intact. The other new and relevant parameters in the model are $g_D$, $g_D^\prime$, $v_D$ and $\lambda_D$, which we would like to constrain by collider physics, astrophysics and cosmology in this work.

\subsection{Collider Physics}

The extension of the scalar sector in the model can be constrained by the data from Higgs search experiments and the measurements of the SM-like Higgs boson at LEP, Tevatron and LHC. 

Here, the Higgs phenomenology at colliders is similar to that of one real singlet scalar extension of the SM. The most relevant parameters are the mixing angle $\alpha$ and the mass of the extra Higgs boson $m_{h_2}$. 
The couplings of the physical scalars $h_1$ and $h_2$ to the SM particles can be given as 
\be
{\cal L}_{\rm Higgs} \supset  \frac{h_1 \cos \alpha - h_2 \sin \alpha}{v_H} \left(2 m_{W}^2 W^{+}_\mu W^{-\mu} + m_Z Z_\mu Z^\mu - \sum_{f} m_f {\bar{f}} f\right) \; . 
\ee
One can see that the SM-like Higgs boson coupling to the SM particles are modified by a factor of $\cos\alpha$. 
The Higgs boson signal strength then can be given as 
\beq
\mu_{h_1} \equiv \cos^2\alpha \frac{{\rm BR}(h_1 \to {\rm SM})}{{\rm BR^{SM}}(h_1 \to {\rm SM})} \; ,
\eeq
where $\text{BR}^{\text{SM}}(h_1\rightarrow \text{SM}) \equiv 1$ and $\text{BR}(h_1\rightarrow \text{SM}) = \frac{ \Gamma_{h_1}^{\text{SM}}\cos^2\alpha}{ \Gamma_{h_1}^{\text{SM}}\cos^2\alpha +\Gamma_{h_1}^{\text{DS}} }$ with $\Gamma_{h_1}^{\text{DS}}$ is the partial decay width of $h_1$ to dark sector particles. Therefore, one obtains 
\beq
\mu_{h_1} = \frac{\Gamma_{h_1}^{\text{SM}}\cos^4\alpha}{\Gamma_{h_1}^{\text{SM}}\cos^2\alpha+\Gamma_{h_1}^{\text{DS}}} \; .
\eeq
It is expected that $\Gamma_{h_1}^{\text{DS}} \ll \Gamma_{h_1}^{\text{SM}}$ thus the signal strength simply scales as $\cos^2\alpha$. Using the current Higgs signal strengths measurement $\mu_h = 1.05 \pm 0.06$ by ATLAS~\cite{ATLAS:2022vkf}, one can obtain a bound on the mixing angle $|\sin\alpha |\lesssim 0.2$ at $95\%$ C.L.

If $m_{h_1} \geq 2 m_{\cal W}$, the Higgs boson $h_1$ can decay invisibly into a pair of ${\cal W}$. The invisible branching ratio can be given by 
\be
{\rm BR} ({h_1 \to {\rm inv}}) =\frac{ \Gamma({h_1 \to {\cal W}^p {\cal W}^{m}})} {\Gamma_{h_1}^{\text{SM}}\cos^2\alpha+\Gamma_{h_1}^{\text{DS}}} \; ,
\ee 
where the invisible decay width can be given as
\be
\Gamma({h_1 \to {\cal W}^p {\cal W}^{m}}) = \frac{g_D^4 \left(v_D \sin \alpha \right)^2 }{256 \pi} 
\frac{m_{h_1}^3}{m_{{\cal W}}^4} \left( 1- \tau_{{\cal W}} +\frac{3}{4}  \tau_{{\cal W}}^2\right) 
\sqrt{1- \tau_{{\cal W}} } \; ,
\label{eq:invdecaywidth}
\ee
with $\tau_{{\cal W}} = 4 m_{{\cal W}}^2 / m_{h_1}^2$. 
Recently, ATLAS sets a limit of ${\rm BR}({h_1 \to{ \rm inv}}) < 0.13$ at $95 \%$ C.L., assuming that the Higgs boson production cross section via vector boson fusion is comparable to the SM prediction~\cite{ATLAS:2022vkf}.

\begin{figure}[tb]
    \centering
    \includegraphics[width=0.7\textwidth]{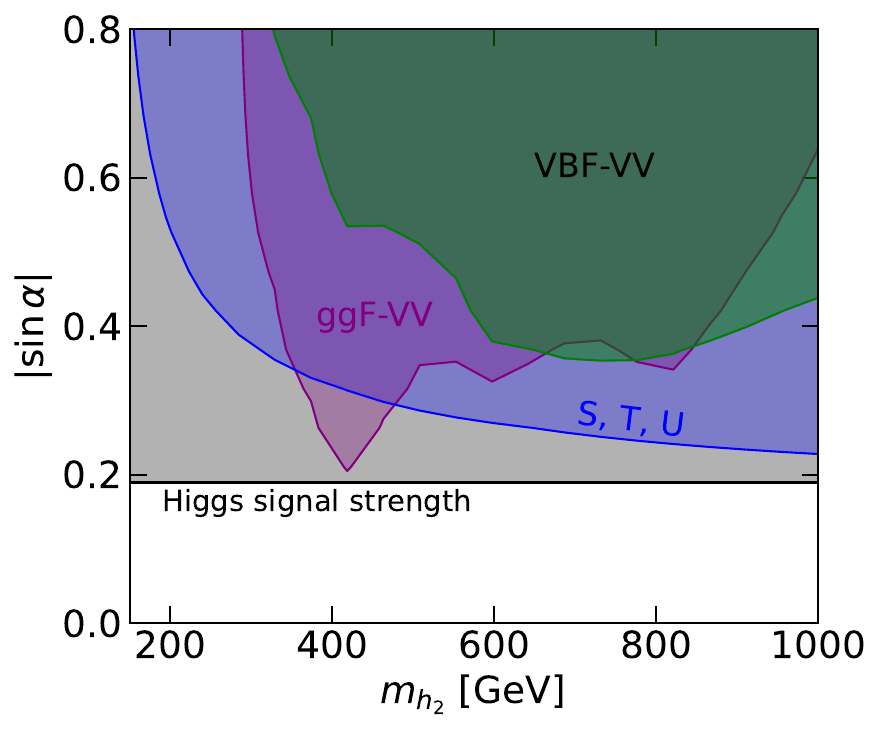}
    \caption{\label{fig:collider}Upper bounds on the mixing angle $\alpha$ as a function of the heavy Higgs mass $m_{h_2}$. The color shaded regions represent the exclusion regions from the Higgs signal strength measurements at ATLAS~\cite{ATLAS:2022vkf} (gray region), the oblique parameter constraint (blue region) and the di-boson searches via gluon-gluon fusion (purple region) and vector boson fusion (green region) channels at ATLAS~\cite{ATLAS:2020fry}. 
    }
\end{figure}

The heavy Higgs boson $h_2$ can interact with the SM particles through the mixing
with the SM-like Higgs boson $h_1$. If kinematically allowed, $h_2$ can decay to a pair of massive gauge bosons, a pair of fermions, a pair of SM-like Higgs bosons or a pair of dark sector particles. 
The resonance production rate of $h_2$ to a pair of SM particles $X$ can be given as 
\be
\sigma(pp \to XX) \equiv \sigma( pp \to h_2) \times {\rm BR} (h_2 \to XX) \; ,
\ee
where $\sigma( pp \to h_2) = \sin^2\alpha \, \sigma^{\rm SM}( pp \to h)|_{m_h = m_{h_2}}$ and 
\be
{\rm BR} (h_2 \to XX) = \frac{ \Gamma (h_2 \to XX)} {\sin^2\alpha \, {\Gamma^{\rm SM}_h}|_{m_h = m_{h_2}} + \Gamma_{h_2}^{\rm DS}} \; .
\ee
The partial decay with of $h_2$ to the dark sector particles $\Gamma_{h_2}^{\rm DS} \simeq \Gamma(h_2 \to Z_2 Z_2) + \Gamma(h_2 \to Z_3 Z_3) + \Gamma(h_2 \to {\cal W}^p {\cal W}^{m})$. 
We utilize the current direct heavy resonance searches at the LHC including $h_2 \to W^+W^-, ZZ$ \cite{ATLAS:2020fry}, $h_2 \to \tau^+ \tau^-$ \cite{ATLAS:2020zms} and $h_2 \to \mu^+\mu^-$ \cite{ATLAS:2019odt} to set constraints on the model parameter space. 
We find that the current data from $\tau^+ \tau^-$ and $\mu^+\mu^-$ final states are unlikely to constraint the model, however the di-boson final states data can put an upper bound on the mixing angle as depicted as purple and green shaded regions in Fig.~\ref{fig:collider}. We assume $\Gamma_{h_2}^{\rm DS} \sim 0$ for the result shown in Fig.~\ref{fig:collider}, and notice that a larger value of this partial decay width leads to a less stringent upper bound on the mixing angle from the heavy Higgs boson searches. 

\subsection{Electroweak Precision Measurements and Dark Photon/$Z^\prime$ Physics}

The presence of the extra scalar $h_2$ and its mixing to the SM-like Higgs boson can modify the oblique parameters $S$, $T$, and $U$ \cite{Peskin:1991sw}. 
In particular, any oblique parameter $\cal O$ in the model can be given as \cite{Profumo:2014opa} 
\be
\Delta {\cal O} = \left[{\cal O}^{\rm SM} (m_{h_2}) - {\cal O}^{\rm SM} (m_{h_1}) \right] \, \sin^2 \alpha \; ,
\ee
where ${\cal O}^{\rm SM}$ is the oblique parameter given in the SM. 
We utilize the global fit values for the oblique parameters taken from the 
Particle Data Group (PDG)~\cite{Zyla:2020zbs},
which are given as 
\bea
\label{eq:STU_pdg}
\Delta S &=& -0.01 \pm 0.1 \; , \nonumber \\
\Delta T &=& 0.03 \pm 0.12 \; ,  \\
\Delta U &=& 0.02 \pm 0.11 \; , \nonumber
\eea 
and the correlation coefficients are $0.92, -0.8$ and $-0.93$ for ($\Delta S, \Delta T$), ($\Delta S, \Delta U$) and ($\Delta T, \Delta U$), respectively. 
The constraint on the mixing angle $\alpha$ versus the heavy Higgs mass from the oblique parameters are shown as the blue shaded region in Fig.~\ref{fig:collider}. One can see that the upper bound on the mixing angle becomes more stringent in the heavier mass region of $h_2$. At $m_{h_2} = 1$ TeV, the mixing angle $|\sin \alpha| \lesssim 0.23$.

The presence of extra neutral gauge bosons and its mixing with the SM gauge boson given in (\ref{MassV3}) can modify the physics at $Z$-pole. 
The most relevant parameter to the modification is the kinetic mixing $\epsilon$. We compute the $Z$-pole physics shifts and use the electroweak precision measurements given in PDG \cite{Zyla:2020zbs} to constraint the model. 

Moreover, via the kinetic mixing, the dark photon and dark $Z^\prime$ can decay to the SM particles if they are kinematically allowed. Their decay widths to the SM leptons can be given as 
\begin{equation}
\label{eq:Apdecay}
\Gamma\left(Z_{i} \to \bar{\ell}\ell\right) = \frac{\alpha}{3}\varepsilon_{\ell}^2 m_{Z_{i}}
\sqrt{1 - \mu_\ell^2}\left(1 + \frac{\mu_\ell^2}{2}\right),
\end{equation}
where $i = \{2,3\}$, $\mu_\ell = 2m_\ell/m_{Z_{i}}<1$ since it only opens for
$m_{Z_{i}} > 2 m_\ell$ and the mixing parameter $\varepsilon_\ell$ is given at tree level by
\begin{equation}
\label{eq:epsilon}
\varepsilon_\ell = \frac{1}{2 g s_W }\sqrt{\left(C_{V}^{\ell i}\right)^2
	+ \left(C_{A}^{\ell i} \right)^2\left(\frac{1 - \mu_\ell^2}{1 + \mu_\ell^2/2}\right)}\;,
\end{equation}
with $C_{V}^{\ell i}$ and $C_{A}^{\ell i}$ are given in (\ref{eq:CV}) and (\ref{eq:CA}), respectively. 
Current searches for the dark photon and dark $Z^\prime$ set constraints on the mixing parameter $\varepsilon_\ell$ and $m_{Z_{2,3}}$. We focus on the mass regions $m_{Z_{2,3}} > 1$ MeV and take into account the current dark photon searches from collider/fixed target experiments including A1~\cite{Merkel:2014avp}, KLOE~\cite{KLOE-2:2011hhj,KLOE-2:2012lii,KLOE-2:2014qxg,KLOE-2:2016ydq},
NA48/2~\cite{NA482:2015wmo}, 
BaBar~\cite{BaBar:2014zli}, 
LHCb~\cite{LHCb:2019vmc}, 
CMS~\cite{CMS:2019buh}   
and ATLAS~\cite{ATLAS:2019erb}, 
and beam dump experiments including 
E774~\cite{Bross:1989mp}, E141~\cite{Riordan:1987aw}, E137~\cite{Bjorken:1988as,Batell:2014mga,Marsicano:2018krp}),
$\nu$-Cal~\cite{Blumlein:2011mv,Blumlein:2013cua},
and CHARM ~\cite{Gninenko:2012eq}, 
as well as the bounds from supernovae~\cite{Chang:2016ntp} and $(g-2)_e $~\cite{Pospelov:2008zw}. 
Recent constraints from FASER~\cite{FASER:2023tle} (decaying to $e^+ e^-$) and NA62~\cite{NA62:2023qyn} (decaying to $\mu^+ \mu^-$) experiments are also taken into account in this analysis. 

\subsection{Relic Density}

The DM candidate ${\cal W}$ can annihilate to SM particles via the $2\to 2$ $s$ channel processes with $h_i$ and $Z_i$ as mediators. In addition, they can annihilate to $Z_i Z_j$ and $h_i h_j$ via seagull, $t$ channel and $u$ channel diagrams. 

Here we utilize {\tt micrOMEGAs} package \cite{Belanger:2018ccd} to calculate the DM relic density and compare against the observed DM relic density, $\Omega_{\rm DM} h^2 = 0.120 \pm 0.001$ from the Planck Collaboration~\cite{Planck:2018vyg}. 

We note that the annihilation process ${\cal W}^p {\cal W}^{m} \to \gamma \gamma$ can be occurred via the SM charged fermion and $W^\pm$ loops (triangle diagrams with $h_i/Z_j$ exchange connected to the $\mathcal W$s) and therefore negligible in the relic abundance computation! It may give rise to discrete gamma ray lines in the sky for indirect detection which is important and deserves a separate study by itself.

\subsection{Direct Detection}

The DM candidate ${\cal W}$ can be directly observable when it scatters off the target material within underground detectors, resulting in a detectable recoil energy. The scattering process between ${\cal W}$ and nucleons can occur through $t$ channel diagrams involving mediators $h_i$ and $Z_i$ in the model. 
The elastic cross section between DM and a nucleon, denoted as $N$, can be expressed as
\begin{align}
\label{eq:DDxsec}
\sigma^{\rm SI}_{{\cal W} N} & = 
	\sigma^{\rm SI}_{{\cal W}p}
	\frac{\sum_k \eta_k \mu_{A_k}^2 \left[Z_\text{atom} + (A_k - Z_\text{atom}) f_n/f_p\right]^2}
		{\sum_k \eta_k \mu_{A_k}^2 A_k^2} \;.
\end{align}
Here $\sigma^{\rm SI}_{{\cal W}p}$ represents the elastic cross section between  DM and a proton, 
$\mu_{p}$ is the reduced DM-proton mass defined as 
$\mu_{p} = m_{{\cal W}} m_p/(m_{{\cal W}} + m_p)$,
$\mu_{A_k} = m_{{\cal W}}m_{A_k}/(m_{{\cal W}} + m_{A_k})$ is the reduced DM-isotope
nucleus mass and $f_p$ and $f_n$ are effective couplings of the DM with
protons and neutrons, respectively.
The atomic number is denoted as $Z_\text{atom}$, while $\eta_k$ and $A_k$ represent isotope-specific variables indicating the abundance and mass number of the $k^\text{th}$ target isotope, respectively.

Typically, direct detection experiments assume isospin conservation, \textit{i.e.}, $f_p = f_n$, and report results in the form of (\ref{eq:DDxsec}). However, in this model, the couplings between the gauge boson mediator particles $Z_{i}$ and quarks, specifically $u$ and $d$, can differ due to their distinct SM charges, leading to isospin violation, where $f_p \neq f_n$.
Following Refs.~\cite{Feng:2011vu,Yaguna:2016bga}, we can rescale the reported experimental
limit, $\sigma_\text{limit} \to \sigma_\text{limit}\times\sigma^{\rm
SI}_{{\cal W}p}/\sigma^{\rm SI}_{{\cal W}N}$ to account for the effects of isospin violation. This rescaled value can be used to constrain $\sigma^{\rm SI}_{{\cal W}p}$. 

We use {\tt micrOMEGAs} package to calculate the effective couplings $f_p$ and $f_n$ as well as the elastic cross section $\sigma^{\rm SI}_{{\cal W}p}$. 
The results of the cross section are compared against recent upper limits from experiments including 
CRESST III~\cite{CRESST:2017ues}, DarkSide-50~\cite{DarkSide:2018bpj,DarkSide-50:2022qzh},
XENON1T \cite{XENON:2018voc, XENON:2019gfn}, 
XENONnT~\cite{XENON:2023cxc}, PandaX-4T~\cite{PandaX-4T:2021bab}, 
and LZ~\cite{LZ:2022lsv}. 

\begin{figure}[tb]
    \centering
    \includegraphics[width=0.4\textwidth]{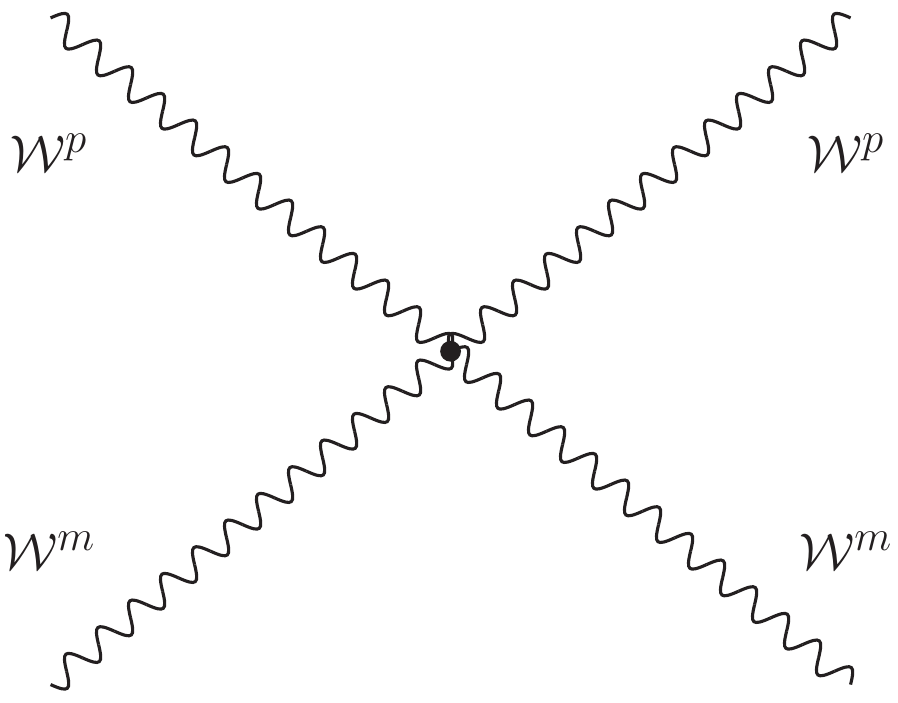}
    \caption{\label{fig:FeynDiag1} The seagull diagram for SIDM in the model.} 
\end{figure}

\begin{figure}[tb]
    \centering
    \includegraphics[width=0.7\textwidth]{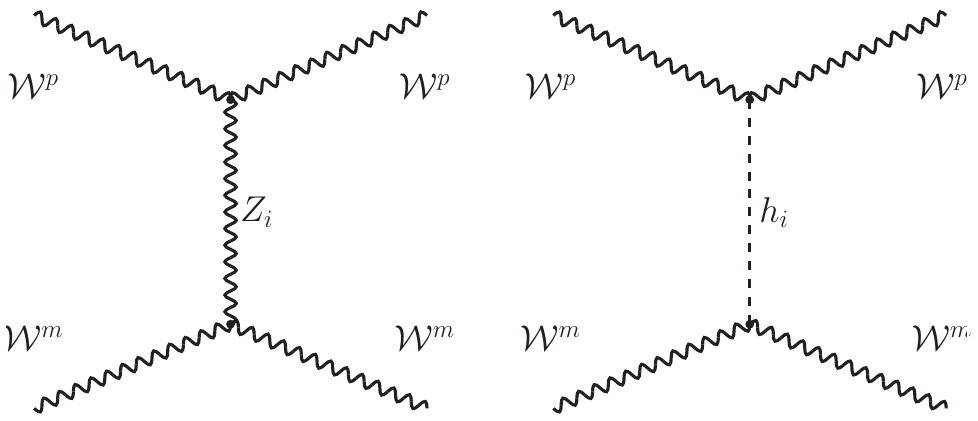}
    \caption{\label{fig:FeynDiag2} The $t$ channel Feynman diagrams for SIDM in the model.} 
\end{figure}

\section{Self Interacting Dark Matter \label{sec:SIDM}}

The non-Abelian nature of gauge bosons enables the self-interaction of DM $\mathcal W$ via the seagull diagram as depicted in Fig.~\ref{fig:FeynDiag1}. 
Furthermore, $\mathcal W$ self-interaction can be facilitated through $t$ channel diagrams involving the $h_i$ and $Z_i$ bosons as mediators, depicted in Fig.~\ref{fig:FeynDiag2}. While the seagull diagram interaction results in a velocity-independent cross section, the mediator exchange diagrams can give rise to long-range interactions with velocity-dependence.
When the mediator involves a vector boson (\textit{i.e.}, $Z_i$ exchange), interactions between the DM particle ${\cal W}^m$ and its antiparticle ${\cal W}^{p}$ result in an attractive potential, while interactions between two identical particles ${\mathcal W}^m$ (or two identical antiparticles ${\cal W}^p$) yield a repulsive potential. Conversely, if the mediator is a scalar boson (\textit{i.e.}, $h_i$ exchange), the DM self-interactions manifest as purely attractive.
Since we are considering a symmetric DM (equal ${\cal W}^m$ and ${\cal W}^p$), it exhibits both attractive and repulsive self-interactions DM and hence the total effective cross section will be given as the average of the two.

The momentum transfer cross section for the DM self interacting is obtained by  
\begin{equation}\label{sigt}
\sigma_{\rm T} \equiv \int d\Omega \, (1-\cos \theta)\frac{d\sigma}{d\Omega} \; ,
\end{equation}
where $\theta$ is the scattering angle and  $(1-\cos \theta)$ is the fractional change in longitudinal momentum. 
The cross section is calculated according to different regimes defined by two dimensionless parameters $\kappa$ and $\beta$ introduced in (\ref{eq:kappaandbeta}) and presented again here for convenience
\beq 
\label{eq:kappaandbeta2}
\kappa = \frac{m_{\cal W} v}{m_{X}} \; , \quad\quad  \beta=\frac{2\alpha_X m_{X}}{m_{\cal W}v^2 } \;, 
\eeq
where $X \equiv \{{h_i}, {Z_i} \}$ in the model and thus
\be
\label{eq:alphaX}
\alpha_{h_1} = \frac{1}{4\pi} \left(\frac{ g_D \sin \alpha }{2} \right)^2 \,,\;\;\;\;
\alpha_{h_2} = \frac{1}{4\pi} \left(\frac{ g_D \cos \alpha }{2} \right)^2 \,,\;\;\;\;
\alpha_{Z_i} = \frac{\left( g_D {\cal O}_{3i} \right)^2}{4 \pi} \,.
\ee
In particular, the momentum transfer cross section can be computed in the Born approximation regime if $2\beta\kappa^2 \ll 1$, in the semi-classical regime if $\kappa\geq1$ 
and in the quantum regime if $\kappa\ll1$. 

In the Born approximation, the cross section can be computed perturbatively in $\alpha_X$. 
On the other hand, in the semi-classical and quantum regimes, the non-perturbative effects become significant. 
We follow the calculation method in \cite{Colquhoun:2020adl} for the semi-classical regime.  
In particular, the momentum transfer cross section in the semi-classical regime for an attractive Yukawa potential is given by 
\begin{align}
\sigma_\mathrm{T}^{\mathrm{att.}} &= \frac{\pi}{m_{ X}^2} \times \begin{cases}
2\beta^2 \zeta_{1/2}\left(\kappa, \beta\right) & \beta\leq 0.2 \,, \\
\hspace{7cm}\  & \  \\[-4mm]
2\beta^2 \zeta_{1/2}\left(\kappa, \beta\right) e^{0.64(\beta - 0.2)} & 0.2 < \beta \leq 1\,, \\
\hspace{7cm}\  & \  \\[-4mm]
4.7 \log(\beta + 0.82)  & 1 < \beta < 50\,, \\ 
\hspace{7cm}\  & \  \\[-4mm]
2 ( \log \beta ) (\log \log \beta + 1) & \beta \geq 50\;,
\end{cases} 
\label{eq:sigmaTatt}
\end{align}
and for a repulsive Yukawa potential,
\begin{align}
\sigma_\mathrm{T}^{\mathrm{rep.}} & = \frac{\pi}{m_{ X}^2} \times \begin{cases}
2\beta^2\zeta_{1/2}\left(\kappa, \beta\right) & \beta\leq0.2\,,\\
\hspace{7cm}\  & \ \\[-4mm]
2\beta^2\zeta_{1/2}\left(\kappa, \beta\right) e^{-0.53(\beta - 0.2)} & 0.2 < \beta \leq 1\,, \\
\hspace{7cm}\  & \ \\[-4mm]
2.9 \log(\beta + 0.47) & 1 < \beta < 50\,, \\
\hspace{7cm}\  & \ \\[-4mm]
\lambda_\mathrm{T} (\log 2\beta - \log \log 2 \beta)^2 & \beta \geq 50\;,
\end{cases} 
\label{eq:sigmaTrep}
\end{align}
where
\begin{align}
\label{zeta}
\lambda_\mathrm{T} & = \frac{1}{2}(1 + \cos 2 + 2 \sin 2) \; , \\
\zeta_n(\kappa,\beta) & =\frac{\text{max}(n,\beta\kappa)^2 - n^2}{2 \kappa^2 \beta^2} + \eta\left(\frac{\text{max}(n,\beta\kappa)}{\kappa}\right)\; ,\\
\eta(x) & = x^2 \left[ -K_1\left(x\right)^2 + K_0\left(x\right) K_2\left(x\right)\right]\;,
\end{align}
with $n = 1$ for distinguishable particles and $n = \tfrac{1}{2}$ ($n = \tfrac{3}{2}$) for identical particles with even (odd) spatial wave function, and $K_n (x)$ being the modified Bessel function of the second kind of order $n$. 

For the quantum regime where $\kappa<0.4$, the cross section is largely dominated by $S$-wave scattering and 
the Schr\"odinger equation can be solved analytically for the Hulth\'en potential \cite{Tulin:2013teo}, 
\begin{equation}
\label{Hulthenl}
\sigma_{T}^{\mbox{Hulth\'en}}=\frac{16\pi}{m_{\cal W}^2v^2}\sin^2\delta_0 \; ,
\end{equation}
where the phase shift $\delta_0$ is given by
\begin{equation}\label{phaseshift}
\delta_0 =\arg\left(i \frac{\Gamma(l_{p} + l_{m} -2)}{\Gamma(l_{p})\Gamma(l_{m})}\right) \; , 
\end{equation}
with
\begin{equation}
\label{eq:lplm}
\begin{split}
l_{p}& = 1 + i \frac{\kappa}{{2\rho}}(1 + \sqrt{1 \pm 2\rho\beta}) \; ,\\
l_{m}& = 1 + i \frac{\kappa}{{2\rho}}(1 - \sqrt{1 \pm 2 \rho \beta}) \; .
\end{split}
\end{equation}
Here $\rho \simeq 1.6$ is a dimensionless number and the signs $+(-)$ denotes repulsive (attractive). 

For the region between semi-classical and quantum regimes, \textit{i.e.} $ 0.4 < \kappa < 1$, we follow \cite{Colquhoun:2020adl} to use the interpolation function  
\begin{eqnarray}
  \sigma_{T}=(1-\kappa) \,\sigma_{T}^{\mbox{Hulth\'en}} /0.6 +(\kappa-0.4) \, \sigma_{ T}^{\rm{rep(att)}} /0.6 \; .
\end{eqnarray}

Finally, to compare against the astrophysical data, we compute the velocity-averaged transfer cross section as~\cite{Colquhoun:2020adl} 
\be 
\label{sigmaTv}
\overline{\sigma_{T}} = \frac{\langle \sigma_T v^2 \rangle}{16 \sqrt{2} v_0^2/\pi}   = \frac{1}{16 \sqrt{2} v_0^2/\pi}  \int_0^{\infty} f(v,\sqrt{2} v_0) \sigma_T v^2 dv \; , 
\ee 
where $v$ is the relative velocity in the center-of-mass frame, $v_0= \sqrt{\pi}\langle v\rangle/4$ is the velocity dispersion and the Maxwell-Boltzmann distribution with parameter $a$ is given by
\be
f(v,a)=\sqrt{\frac{2}{\pi} } \frac{v^2 e^{-v^2/(2 a^2)}}{a^3} \; .
\ee

\begin{figure}[tb]
    \centering
    \includegraphics[width=0.7\textwidth]{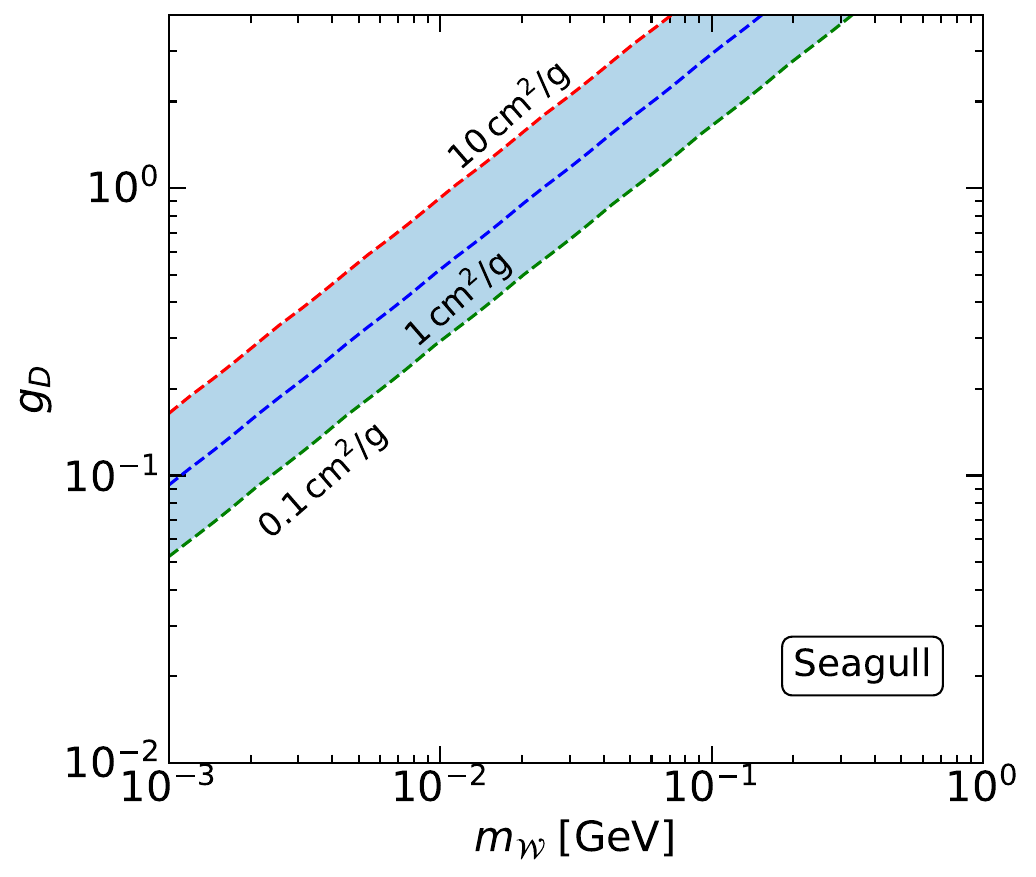}
    \caption{\label{fig:seagull}The momentum transfer cross section divided the DM mass from the seagull diagram spanned in ($m_{\cal W}, g_D$) plane. The blue-shaded region is favored for the astrophysical observations in dwarfs, LSB spiral galaxies and galaxy clusters. The dashed green, blue and red lines represent $\sigma_T/{m_{\cal W}} = 0.1, 1$ and $10 \, \rm{cm}^2/g$ respectively.} 
\end{figure}

\section{Numerical Results \label{numbers}}

We begin by showing in Fig.~\ref{fig:seagull} the Born cross section solely from the seagull diagram spanned in the ($m_{\cal W}, g_D$) plane. The blue band indicates the parameter space where $\sigma_{T}/m_{\cal W} = 0.1 - 10 \, \rm{cm}^2/g$, as preferred for solving small scale issues. 
The dark matter mass $m_{\cal W}$ lies in sub-GeV scale for the new gauge coupling $g_D$ of order unity.

Next, we explore the parameter space of the model that 
satisfies all the theoretical and experimental constraints discussed in the previous sections. In particular, we include the theoretical constraints on the scalar potential, the Higgs measurements, the direct searches for heavy Higgs, the electroweak precision measurements, dark photon, dark $Z^\prime$ searches and the dark matter relic density as well as the dark matter direct detection. 
We compute the self interacting DM cross section as discussed in section \ref{sec:SIDM} and then focus on the parameter space that give $ 0.1 \, {\rm{cm}^2/g } < \sigma_{T}/m_{\cal W} \, (\langle v\rangle = 10 \, {\rm km/s }) < 10 \, \rm{cm}^2/g$ as preferred for solving the small scale issues.
To sample the parameter space in the model, we employ MCMC scans using {\tt emcee}~\cite{Foreman-Mackey:2012any}.
The parameter range is set as follows, 
\bea
m_{h_2}/ {\rm GeV} &\in& [150,\, 2000 ]  \; , \\
\alpha/ {\rm rad}  &\in& \left[-0.2,\, 0.2 \right] \; , \\ 
m_{{\cal W}}/{\rm GeV}&\in& [10^{-3},\, 10^3] \; , \\
M_{X}/{\rm GeV} &\in& [ 10^{-3},\, 80 ]   \; , \\
g_D  &\in&  [ 10^{-6},\, 1]  \; , \\
g_D'  &\in& [10^{-6},\, 1 ] \; , \\
\epsilon  &\in& [ 10^{-15},\, 10^{-1} ] \; .
\eea
Here $m_{h_2}$ and $\alpha$ are sampled in the linear scale, while the remaining free parameters are in the log scale prior. 
With the above choice of scan parameters, 
$v_D$ is given by $2 m_{\mathcal W}/g_D$, while $\lambda$,
$\lambda_D$ and $\lambda^\prime$ are computed by using (\ref{lambda}), (\ref{lambdaD}) and (\ref{lambdaprime}) respectively.
We note that the lower scanning range for $\epsilon$ is set to a very small value due to the stringent constraint imposed on the dark photon within the low mass region of $Z_{2,3}$, where $m_{Z_{2,3}} \sim {\cal O}(100 \, {\rm MeV})$, as per the 
E137~\cite{Bjorken:1988as,Batell:2014mga,Marsicano:2018krp}, $\nu$-Cal~\cite{Blumlein:2011mv,Blumlein:2013cua} and CHARM ~\cite{Gninenko:2012eq} experiments as well as the supernova bound \cite{Chang:2016ntp}.

\begin{figure}[tb]
    \centering
    \includegraphics[width=0.7\textwidth]{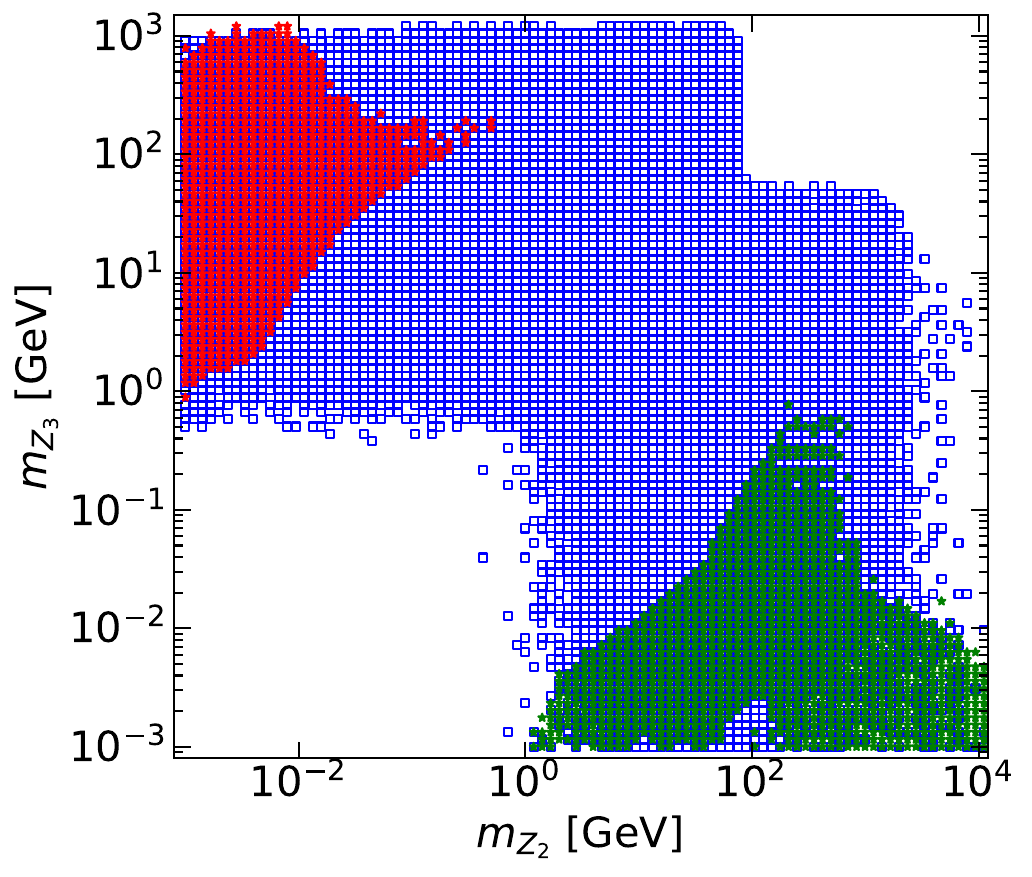}
    \caption{\label{fig:numerical_res0}The $2\sigma$ favored region projected on the ($m_{Z_2}$, $m_{Z_3}$) plane. The blue boxes indicate the favored region considering all model constraints except the SIDM, while the red and green stars represent the inclusion of SIDM with the dominance of $Z_2$ and $Z_3$ mediation diagrams, respectively.} 
\end{figure}

Fig.~\ref{fig:numerical_res0} shows the $2\sigma$ favored region ($\Delta \chi^2 < 5.99$) projected on ($m_{Z_2}$, $m_{Z_3}$) plane. To understand the impact of the SIDM, we illustrate two superimposed regions: one taking into account the SIDM constraint (regions marked by red and green stars) and another that does not include the SIDM constraint (blue boxed region).
We find that the dominant contribution to the SIDM cross section arises from either the $Z_2$ or $Z_3$ mediation diagram, depending on their respective mass spectra, with the lighter mass one significantly influencing the magnitude of this contribution. 
Upon scrutinizing the constraints imposed by astrophysical data on SIDM cross section, discrete and isolated regions are discernible within the parameter space, each indicative of distinctive primary contributions. In particular, the region marked by red stars, where $m_{Z_2}$ < 400 MeV and $m_{Z_3} > 1$ GeV, corresponds to the $Z_2$ mediation diagram dominant region while the green stars region where  $m_{Z_2} > 1$ GeV and $m_{Z_3} < 600$ MeV, represents the region that the $Z_3$ mediation diagram dominant. Following we will show the results for the mass order relation $m_{Z_2} < m_{Z_3}$, thereby focusing exclusively on the region where the cross section is dominated by the $Z_2$ mediation diagram. The results for the reverse mass order relation $m_{Z_2} > m_{Z_3}$ can be found in the appendix~\ref{app:result_case2}. 

\begin{figure}[tb]
    \centering
    \includegraphics[width=0.49\textwidth]{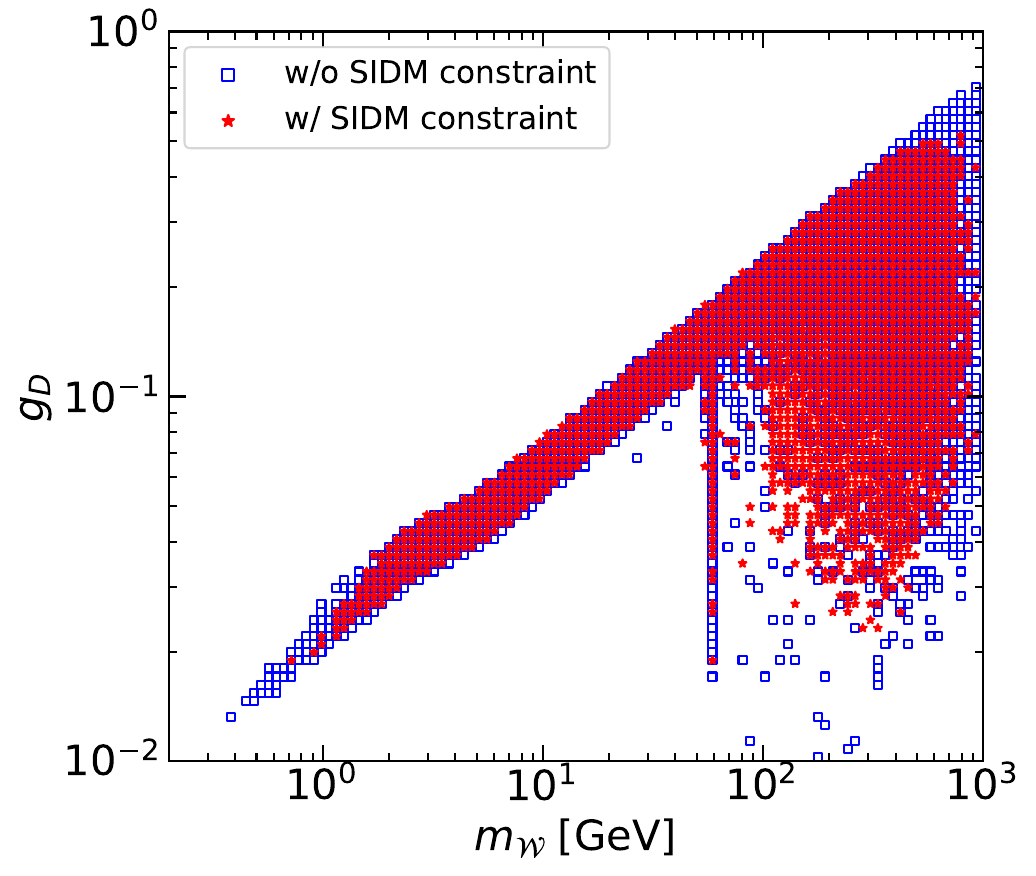}
    \includegraphics[width=0.49\textwidth]{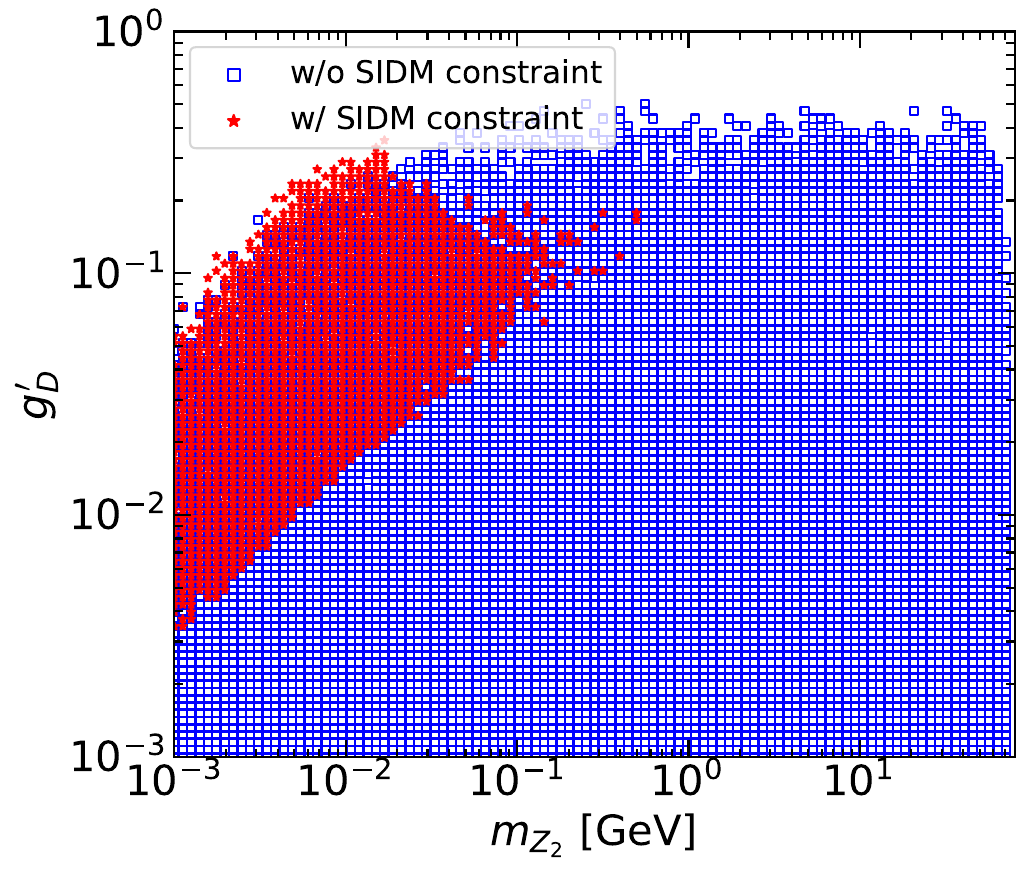}
    \caption{\label{fig:numerical_res1}The $2\sigma$ favored region projected on the ($m_{\cal W}$, $g_D$) plane (left panel) and the ($m_{Z_2}$, $g_D'$) plane (right panel). The blue boxes indicate the favored regions considering all model constraints except the SIDM, while the red stars represent the inclusion of SIDM.} 
\end{figure}

Fig.~\ref{fig:numerical_res1} shows the favored region projected on 2D parameter space of the model. Here, we again illustrate two superimposed regions as in Fig.~\ref{fig:numerical_res0} but by assuming $m_{Z_2} < m_{Z_3}$, thus the regions marked by green stars disappeared. 
In the left panel of Fig.~\ref{fig:numerical_res1}, the favored region is projected on the ($m_{\cal W}$, $g_D$) plane. 
The constraints observed on the upper and lower bounds of the coupling parameter $g_D$ at a certain DM mass primarily arise from the necessity to meet the DM relic density criteria. A larger value of $g_D$ results in an increased DM annihilation cross section, consequently leading to an underabundance of DM relic, while a decreased value (except for the DM annihilation resonance regions) causes an overabundance of DM relics.
Notably, the discernible dip occurring around $m_{\cal W} \simeq 63$ GeV can be attributed to the resonance facilitated by the SM-like Higgs in the DM annihilation process. Additionally, resonance processes involving the heavy Higgs $h_2$ are evident at $m_{\cal W} \geq 75$ GeV, particularly within the realm of small $g_D$ couplings.
We note that since we scan the $h_2$ mass in the heavy region, the DM annihilation resonance process via $h_2$ exchange does not occur in lower DM mass regions.

Furthermore, it becomes apparent that the constraint imposed by SIDM establishes not only upper and lower bounds for $g_D$ but also imposes a lower bound on the DM mass itself.
Specifically, in the absence of the SIDM constraint, the viable range for the coupling $g_D$ extends in $[0.01, 0.8]$ and $m_{\cal W} > 0.4$ GeV. However, upon the inclusion of the SIDM constraint, the range for $g_D$ falls within $[0.02, 0.6]$ and $m_{\cal W} > 0.7$ GeV.

In the right panel of Fig.~\ref{fig:numerical_res1}, the favored region is projected on the ($m_{Z_2}$, $g_D'$) plane.
The influence of the SIDM constraint on this parameter space is of considerable significance, demanding a notably substantial $U(1)_X$ gauge coupling, $g_D'$ and the mass region for $Z_2$. Specifically, the imposition of the SIDM constraint leads to the requirement of a relatively elevated range for $g_D'$, falling between $3\times 10^{-3}$ and $0.4$, and constrains the $m_{Z_2}$ values to below $400$ MeV.
This is due to the dominance of the SIDM cross section by the $Z_2$ mediated process. To satisfy a large cross section preferred for solving small scale issues, this process necessitates a relatively sizable dark fine-structure coupling, $\alpha_{Z_2}$ (as defined in (\ref{eq:alphaX})), and presumes a discernible mass hierarchy between the dark matter particle and its mediator as previously studied in literature~\cite{Tulin:2013teo, Kaplinghat:2015aga}.

We note that $Z_2$ can play the role of a dark photon that couples to the SM particles through kinetic mixing. 
Within the $Z_2$ mass allowing by the SIDM constraint ({\it i.e.} $m_{Z_2} < 400$ MeV), the constraints derived from beam-dump experiments and supernova observations require the kinetic mixing parameter to remain diminutive, {\it e.g.} $\epsilon \lesssim 10^{-7}$ at $m_{Z_2} = 300$ MeV. Consequently, this results in a notably weak interconnection between the dark and SM sectors facilitated primarily through exchanges involving the neutral gauge boson $Z_i$.

\begin{figure}[tb]
    \centering
    \includegraphics[width=0.7\textwidth]{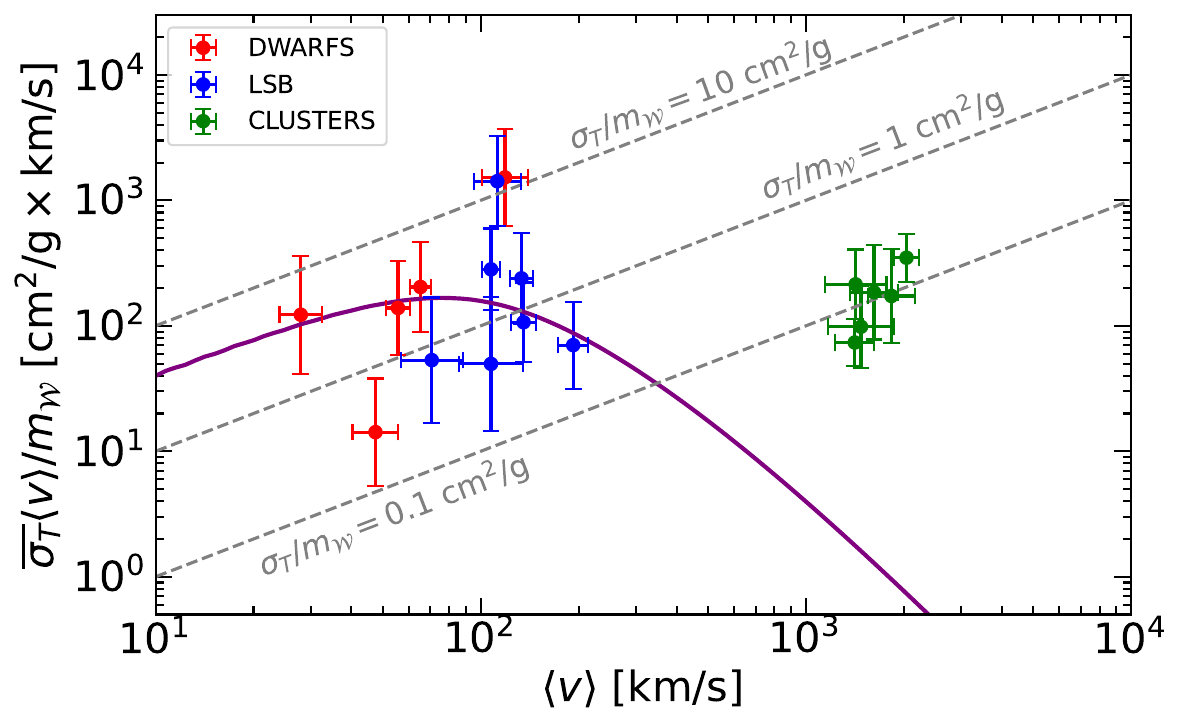}
    \caption{\label{fig:numerical_BM}The SIDM cross section as a function of the average velocity. 
    The red, blue, and green data points represent the astrophysical data from dwarf galaxies, LSB spiral galaxies, and galaxy clusters, respectively. The solid purple line corresponds to the best fit from our model parameter space to the dwarf galaxies and LSB spiral galaxies data. The dashed gray lines indicate the velocity-independent cross sections.} 
\end{figure}

Fig.~\ref{fig:numerical_BM} illustrates the fitting of the SIDM cross section from the model to the astrophysical data sourced from dwarf galaxies, LSB spiral galaxies and galaxy clusters. The astrophysical data are taken from \cite{Kaplinghat:2015aga}. 
The parameter set in the model, which satisfies all current constraints and generates the best fit curve (depicted as the solid purple line in Fig.~\ref{fig:numerical_BM}) for dwarf galaxies and LSB spiral galaxies, is characterized by specific values: $m_{\cal W} = 6.1$ GeV, $M_{\mathcal X} = 6.3$ MeV, $g_D = 0.0446$, $g_D' = 0.0423$, $\epsilon = 2.8 \times 10^{-14}$, $\alpha = -0.034$ radian, and $m_{h_2} = 790$ GeV.
However, extending this fitting to incorporate high-velocity data from galaxy clusters poses a challenge for the model, particularly in explaining the DM abundance from thermal freeze-out. The SIDM cross section required to fit the high-velocity data demands a smaller dark matter mass and relatively large coupling $g_D$. Consequently, this leads to a lower DM relic density in comparison to the observed data from Planck satellite. 

\begin{figure}[tb]
    \centering
    \includegraphics[width=0.8\textwidth]{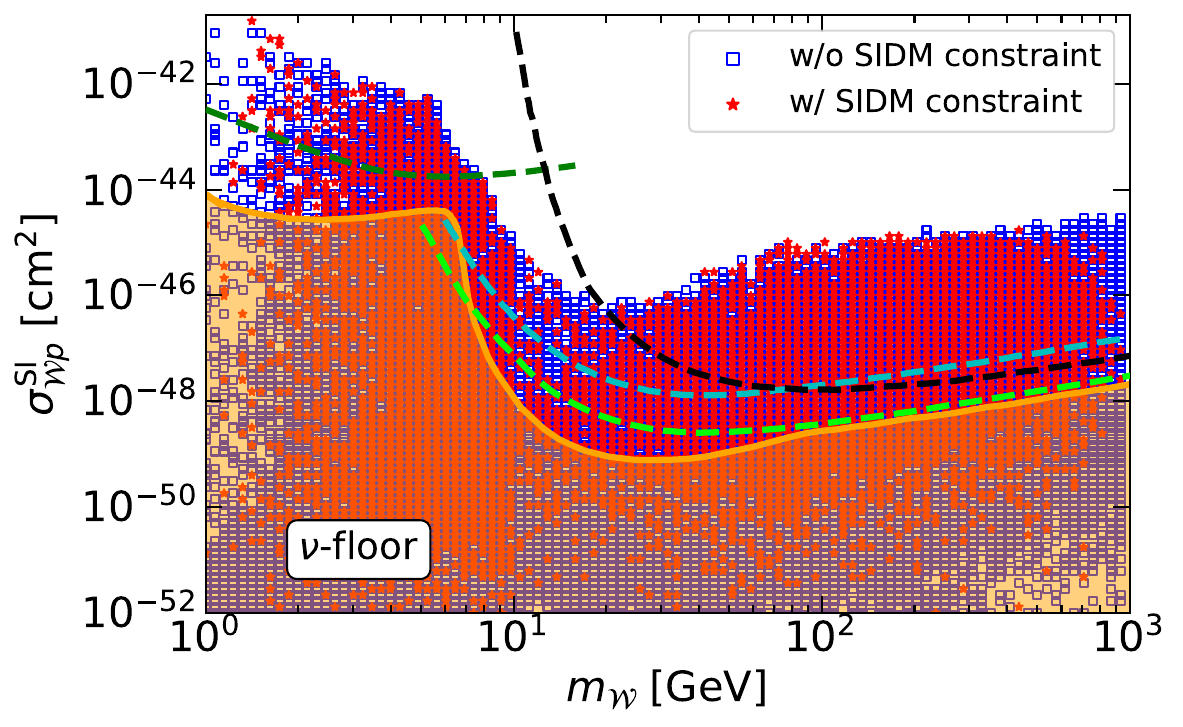}
    \caption{\label{fig:numerical_res2}The scattering cross section between DM and proton as a function of the DM mass. The color scheme corresponds to the encoding used in Fig.~\ref{fig:numerical_res1}. The dashed green, black, cyan and lime lines represent the future sensitivities from CDEX \cite{CDEX:2017kys}, XENONnT \cite{XENON:2020kmp}, DarkSide-20k \cite{DarkSide-20k:2017zyg}, and DARWIN \cite{DARWIN:2016hyl} detectors. The orange region indicates the neutrino floor area. 
    } 
\end{figure}

Fig.~\ref{fig:numerical_res2} shows the favored region projected on the DM-proton spin-independent cross section $\sigma^{\rm SI}_{{\cal W}p}$ and DM mass plane. 
We find that the contributions to the DM-proton scattering cross section are mainly from the diagrams mediated by the Higgs boson $h_1$, and $Z_2$ depending on their masses and couplings to both the DM and the SM particles. 
It is noteworthy that although the cross section $\sigma^{\rm SI}_{{\cal W}p}$ mediated by $Z_2$ can be suppressed due to the smallness of the kinetic mixing parameter $\epsilon$, its effectiveness can be enhanced within the lower mass spectrum of $Z_2$.
Present constraints derived from the direct detection of DM impose an upper limit on $\sigma^{\rm SI}_{{\cal W}p}$. Interestingly, it is important to observe that this limit can appear less stringent in comparison to the one reported by experimental papers, owing to the consideration of isospin violation effects in our analysis.
Of particular interest is the potential to explore the favored region through forthcoming DM direct detection experiments such as CDEX \cite{CDEX:2017kys}, 
XENONnT \cite{XENON:2020kmp}, 
DarkSide-20k \cite{DarkSide-20k:2017zyg} 
and DARWIN \cite{DARWIN:2016hyl}. However, it is crucial to note that a portion of this favored region aligns with the neutrino floor area, where distinguishing dark matter signals from neutrino signals becomes notably challenging. 
Moreover, to distinguish the favored regions with or without the SIDM constraint, one has to go deep
inside the neutrino floor.

\section{Conclusions \label{conclusions}}
In this study, we have conducted a thorough analysis of a novel physics model that extends the SM through the incorporation of a dark version of the SM gauge-Higgs sector. This model introduces a new gauge group, $SU(2)_D \times U(1)_D$, along with the scalar $\Phi_D$, a doublet under $SU(2)_D$ and a singlet under the SM $SU(2)_L \times U(1)_Y$. The interactions and mass spectra of particles in the model were derived, revealing the two mass mixing effects between the dark scalar and the SM-like Higgs as well as between the three neutral gauge bosons. In addition, kinetic mixing between the two factors of abelian $U(1)$s and a Stueckelberg mass of the hidden $U(1)_D$ are included. These setup give rise to three different portals connecting the SM and dark sectors. Notably, the complex vector gauge boson in the dark sector, as demonstrated in this work, can be a vectorial self-interacting dark matter candidate.

Our investigation of the model incorporated both theoretical and current experimental constraints. Theoretical constraints included adherence to perturbative limits for couplings, perturbative unitarity, and the bounded-from-below condition on the scalar potential. Collider phenomenological constraints for the scalar sector were explored using current SM 125 GeV Higgs data and direct searches for heavy Higgs at the LHC. The analysis found that the most stringent upper bound on the mixing angle $|\sin\alpha|$ is set by the 125 GeV Higgs signal strength data from ATLAS, limiting it to less than 0.2. 
The current dark photon/dark $Z^\prime$ searches from collider and beam-dump experiments along with astrophysical observations have also taken into account in this analysis. These constraints limit the kinetic mixing parameter to be tiny, especially in the dark photon/dark $Z^\prime$ mass below $\cal{O}$(100) MeV. 
The electroweak precision measurements on the oblique parameter and $Z$-pole physics were also taken into account properly.

We analyzed the vectorial dark matter phenomenology in the model. 
Dark matter, through Higgs and neutral gauge boson portals, exhibits annihilation to SM particles, and, interestingly, to particles within the dark sector, such as dark photon/dark $Z^\prime$ or dark Higgs, which subsequently decay to SM particles. We found that the latter annihilation channel dominates when the final state particles in the dark sector are light. Moreover, the resonance annihilation process can be occurred and become dominant when the dark matter is about a half of the mediator mass. 
We scanned over the parameter space and computed the thermal relic density of the vector dark matter. Comparing against the observed data from the Planck Collaboration, a linear relationship between the dark matter mass and the $SU(2)_D$ gauge coupling is established in non-resonance annihilation regions. 
A larger dark matter mass requires a correspondingly larger gauge coupling, and at specific mass values, the gauge coupling must fall within a certain range to align with observed data on dark matter relic density. The vectorial dark matter mass ranges from sub-GeV to TeV scale, with the $SU(2)_D$ gauge coupling varying between $\sim 10^{-2}$ and $0.7$. 

The spin-independent cross section for the scattering process between dark matter and nucleons in underground detectors was computed, accounting for isospin violation effects. The primary contributions to the scattering cross section arise from Higgs boson and dark photon exchange diagrams. To assess the validity of our model, we incorporated the latest constraints from dark matter direct detection experiments, including CRESST-III, DarkSide-50, XENON1T, XENONnT, PandaX-4T, and LZ.

Our investigation extended to the self-interaction of vectorial dark matter within the model, with the aim of addressing the small scale ``crisis'' observed in astrophysical phenomena. Self-interaction in the model can occur through the seagull contact interaction and exchange diagrams mediated by dark Higgs and neutral gauge bosons. Notably, with an $SU(2)_D$ gauge coupling of order unity and a dark matter mass around sub-GeV, the former yields a cross section that aligns with resolving the small scale issues. However, this contact interaction results in a velocity-independent cross section, posing challenges in explaining observational data from astrophysical objects spread over vast scales.

On the other hand, the mediator exchange interactions can give rise a velocity-dependent cross section. Our findings indicate that with dark photon/dark $Z'$ boson masses below ${\cal O}$(100) MeV and new gauge couplings ranging from ${\cal O} (10^{-2})$ to ${\cal O}(10^{-1})$, the cross section falls within the expected range to address the 
small scale ``crisis''. Model parameters were fitted to astrophysical observations for dwarf galaxies, low surface brightness spiral galaxies, and galaxy clusters. While the model aligns well with low-velocity data from dwarf and low surface brightness spiral galaxies, reconciling it with high-velocity data from galaxy clusters poses a challenge, particularly when requiring the vectorial dark matter to achieve the correct relic density from Planck data.

Interestingly, good portion of the favorable region of parameter space consistent with all constraints, whether the vectorial dark matter has self-interaction or not, is above the neutrino floor and therefore can be probed by upcoming dark matter direct detection experiments such as CDEX, XENONnT, DarkSide-20k, and DARWIN. This underscores the potential for experimental validation and refinement of our model in light of evolving observational capabilities.

\vskip .5in

\section*{Acknowledgments}
We would like to thank Professor Hai-Bo Yu for useful correspondence and Dr. Sudhakantha Girmohanta for enlightening discussions.
Our special thanks go to Professor Tim Tait for useful suggestions.
This work was supported in part by the NSTC grant No. 111-2112-M-001-035 (TCY) and the National Natural Science Foundation of China, grant No. 19Z103010239 (VQT).
VQT would like to thank the Medium and High Energy Group at the Institute of Physics, Academia Sinica, Taiwan for their hospitality during the progress of this work.

\newpage

\appendix

\section{Neutral Current Interactions \label{app:NCI}}

In this appendix, we present the relevant neutral current interaction Lagrangians for both the fermions and gauge bosons in the model.
Our conventions follow Peskin and Schroeder.~\footnote{M.~E.~Peskin and D.~V.~Schroeder, ``An Introduction to quantum field theory,''
Addison-Wesley, 1995, ISBN 978-0-201-50397-5.}

\subsection{Fermions}

We can express the covariant derivative acting onto any field multiplet $\psi$ in the model in terms of the physical fields $A$ and $Z_i (i=1,2,3)$.
Recall that in general we have~\footnote{
Here $T^\pm = T^1 \pm i T^2$ and $T^3$ (and similarly $\mathcal T^\pm = \mathcal T^1 \pm i \mathcal T^2$ and $\mathcal T^3$) 
where $T^i$ (and $\mathcal T^i$) $(i=1,2,3)$ are the representation of the generators of $SU(2)$ depending on the field multiplet $\psi$;
$Y$ and $X$ are respectively the $U(1)_Y$ hypercharge and $U(1)_D$ dark hypercharge of $\psi$.} 
\begin{align}\label{ConvDGeneral}
D_\mu  =  \partial_\mu & - i \frac{g}{2} \left( W^+_\mu T^+ + W^-_\mu T^- \right) - i g W_\mu^3 T^3 - i g^\prime Y B_\mu  \nn \\
 & - i \frac{g_D}{2} \left( {\mathcal W}^p_\mu {\mathcal T}^p + {\mathcal W}^m_\mu {\mathcal T}^m \right) - i g_D {\mathcal W}^3_\mu \mathcal T^3 - i g_D^\prime X 
{\mathcal X}_\mu   \; .
\end{align}
After the three consecutive transformations of $K$, $O_W$ and $\mathcal O$ obtained above, 
the 4 original neutral fields $B$, $W^3$, $\mathcal X$ and $\mathcal W^3$ transform into physical fields $A$, $Z_1$, $Z_2$ and $Z_3$.
Thus one recasts the covariant derivative in (\ref{ConvDGeneral}) into
\begin{align}
\label{ConvDGeneralPhy}
D_\mu  =  \partial_\mu & - i e Q A_\mu  - i \frac{g}{2} \left( W^+_\mu T^+ + W^-_\mu T^- \right) - i \frac{g_D}{2} \left( {\mathcal W}^p_\mu {\mathcal T}^p + {\mathcal W}^m_\mu {\mathcal T}^m \right)  \nn \\
 &  - i \sum_{i=1}^3 \left[ \frac{g}{c_W} \left(T^3  - s_W^2 Q \right) \mathcal O_{1i} - \left( g^\prime S_\epsilon Y - g^\prime_D C_\epsilon X \right) \mathcal O_{2i}
+ g_D \mathcal T^3 \mathcal O_{3i} \right] Z_i  \; ,
\end{align}
where $Q = T^3 + Y$ is the familiar electric charge operator in SM in unit of $e=g s_W=g^\prime c_W$. 
Using (\ref{ConvDGeneralPhy}), one can work out the couplings of the SM fermions $\psi_L$ and $\psi_R$ with all the gauge fields.
Note that for the SM fermions their $X$ and $\mathcal T^i (i =1,2,3)$ are absent!

Explicitly, the neutral current couplings for the SM fermions are given by
\beq\label{LNC}
\mathcal L_{\rm NC} \supset \sum_i \sum_{f}  \bar f \gamma_\mu \left( C^{fi}_{L}  P_L  + C^{fi}_{R} P_R \right) f Z^\mu_i \; ,
\eeq
where the summation is over all SM fermions $f$ and the three neutral gauge bosons $Z_i$.
The couplings $C^{fi}_{L}$ and $C^{fi}_{R}$ are given by
\begin{align}
C^{ui}_{L} & =  \frac{g}{c_W} \left(\frac{1}{2} - \frac{2}{3} s_W^2 \right) \mathcal O_{1i} - \frac{1}{6} g^\prime S_\delta \mathcal O_{2i} \;, \\
C^{ui}_{R} & =   \frac{g}{c_W} \left(- \frac{2}{3} s_W^2 \right) \mathcal O_{1i} - \frac{2}{3} g^\prime S_\delta \mathcal O_{2i} \;, \\
C^{di}_{L} & =   \frac{g}{c_W} \left( - \frac{1}{2} + \frac{1}{3} s_W^2 \right) \mathcal O_{1i} - \frac{1}{6} g^\prime S_\delta \mathcal O_{2i} \;, \\
C^{di}_{R} & =   \frac{g}{c_W} \left( \frac{1}{3} s_W^2 \right) \mathcal O_{1i} + \frac{1}{3} g^\prime S_\delta \mathcal O_{2i}  \;, \\
C^{\nu i}_{L} & =   \frac{g}{c_W} \left( \frac{1}{2} \right) \mathcal O_{1i} + \frac{1}{2} g^\prime S_\delta \mathcal O_{2i} \; ,\\
C^{\nu i}_{R} & =  \; 0 \;, \\
C^{ei}_{L} & =   \frac{g}{c_W} \left( - \frac{1}{2} + s_W^2 \right) \mathcal O_{1i} + \frac{1}{2} g^\prime S_\delta \mathcal O_{2i}  \; ,\\
C^{ei}_{R} & =  \frac{g}{c_W}  s_W^2 \mathcal O_{1i} +  g^\prime S_\delta \mathcal O_{2i}   \; .
\end{align}

Often $\mathcal L_{\rm NC}$ in (\ref{LNC}) is written in an alternative form in the literature as 
\beq\label{LNC2}
\mathcal L_{\rm NC} \supset \frac{1}{2} \sum_i \sum_{f}  \bar f \gamma_\mu \left( C^{fi}_{V}   - C^{fi}_{A}  \gamma_5 \right) f Z^\mu_i \; ,
\eeq
with 
\begin{align}
\label{eq:CV}
C^{fi}_V & =  C^{fi}_L + C^{fi}_R \; , \\
\label{eq:CA}
C^{fi}_A & =  C^{fi}_L - C^{fi}_R \; .
\end{align}
The couplings presented here are used to obtain the isospin violation effects in direct detection.

\subsection{Gauge Bosons}

The cubic coupling between $W^3$ and $W^+ W^-$ in the SM is given by 
\begin{align}
\mathcal L_{\rm NC}  \supset i g & 
\biggl[
\left( \partial_\mu  W^+_\nu - \partial_\nu W^+_\mu \right) W^{- \mu} W^{3 \, \nu}  - 
\left( \partial_\mu  W^-_\nu - \partial_\nu  W^-_\mu \right)  W^{+ \mu} W^{3 \, \nu} \biggr.  \nn \\
& \biggl. + \frac{1}{2}\left( \partial_\mu W^3_{\nu} - \partial_\nu W^3_{\mu} \right) \left(  W^{+ \mu}  W^{- \nu} -   W^{- \mu} W^{+ \nu} \right) 
\biggr] \; .
\end{align}
Using $W^3 = s_W A + c_W W^{3 \, \prime\prime} = s_W A + c_W \mathcal O_{1i} Z_i$, we obtain the same SM $\gamma W^+ W^-$ coupling
\begin{align}\label{V_gammaWW}
\mathcal L_{\rm NC}  \supset i e & 
\biggl[
\left( \partial_\mu  W^+_\nu - \partial_\nu W^+_\mu \right) W^{- \mu} A^{ \nu}  - 
\left( \partial_\mu  W^-_\nu - \partial_\nu  W^-_\mu \right)  W^{+ \mu} A^{ \nu} \biggr.  \nn \\
& \biggl. + \frac{1}{2}\left( \partial_\mu A_{\nu} - \partial_\nu A_{\mu} \right) \left(  W^{+ \mu}  W^{- \nu} -   W^{- \mu} W^{+ \nu} \right) 
\biggr] \; ,
\end{align}
where we have used the well-known relation $e = g s_W$. 
The $Z_i W^+ W^-$ is given by 
\begin{align}
\mathcal L_{\rm NC}  \supset i g \sum_{i=1}^3 \mathcal O_{1i} & 
\biggl[
\left( \partial_\mu  W^+_\nu - \partial_\nu W^+_\mu \right) W^{- \mu} Z^{\nu}_i  - 
\left( \partial_\mu  W^-_\nu - \partial_\nu  W^-_\mu \right)  W^{+ \mu} Z^{\nu}_i \biggr.  \nn \\
& \biggl. + \frac{1}{2}\left( \partial_\mu Z_{i \, \nu} - \partial_\nu Z_{i \, \mu} \right) \left(  W^{+ \mu}  W^{- \nu} -   W^{- \mu} W^{+ \nu} \right) 
\biggr] \; .
\end{align}
Similarly, the cubic coupling between $Z_i$ and $\mathcal W^p \mathcal W^m$ is given by
\begin{align}
\mathcal L_{\rm NC}  \supset i g_D \sum_{i=1}^3 \mathcal O_{3i} & 
\biggl[
\left( \partial_\mu \mathcal W^p_\nu - \partial_\nu \mathcal W^p_\mu \right) \mathcal W^{m \mu} Z^{\nu}_i  - 
\left( \partial_\mu \mathcal W^m_\nu - \partial_\nu \mathcal W^m_\mu \right) \mathcal W^{p \mu} Z^{\nu}_i \biggr.  \nn \\
& \biggl. + \frac{1}{2}\left( \partial_\mu Z_{i\nu} - \partial_\nu Z_{i\mu} \right) \left( \mathcal W^{p \mu} \mathcal W^{m \nu} -  \mathcal W^{m \mu} \mathcal W^{p \nu} \right) 
\biggr] \; .
\end{align}

Recall that the SM Feynman rule for the $\gamma(p_3,\lambda) W^+(p_2,\nu)W^-(p_1,\mu)$ 
vertex is 
\beq
\label{gammaWWFeynmanRule}
-ie\left[ 
\left(p_1-p_2\right)_\lambda g_{\mu\nu} +
\left(p_2-p_3\right)_\mu g_{\nu\lambda}  +
\left(p_3-p_1\right)_\nu g_{\lambda\mu} 
\right] \; ,
\eeq
with all momenta flowing into the vertex.
Thus the Feynman rules for the $Z_i  W^+  W^-$ and $Z_i \mathcal W^p \mathcal W^m$ vertices can be  deduced simply
 by substituting $e$ in (\ref{gammaWWFeynmanRule}) with $g \mathcal O_{1i}$ and $g_D \mathcal O_{3i}$ respectively. The mixing matrix element $\mathcal{O}_{3i}$ connects the DM $\mathcal W^{p,m}$ with the SM particles via the $Z_i$ exchange.
Note that there is no $\gamma \mathcal W^p \mathcal W^m$ since $W^3$ doesn't mix with $\mathcal W^3$. $\mathcal W^{p,m}$ carry only dark charge but no electric charge.

\newpage

\section{The Reverse Mass Ordering Case of $m_{Z_2} > m_{Z_3}$\label{app:result_case2}}

\begin{figure}[t]
    \centering
    \includegraphics[width=0.49\textwidth]{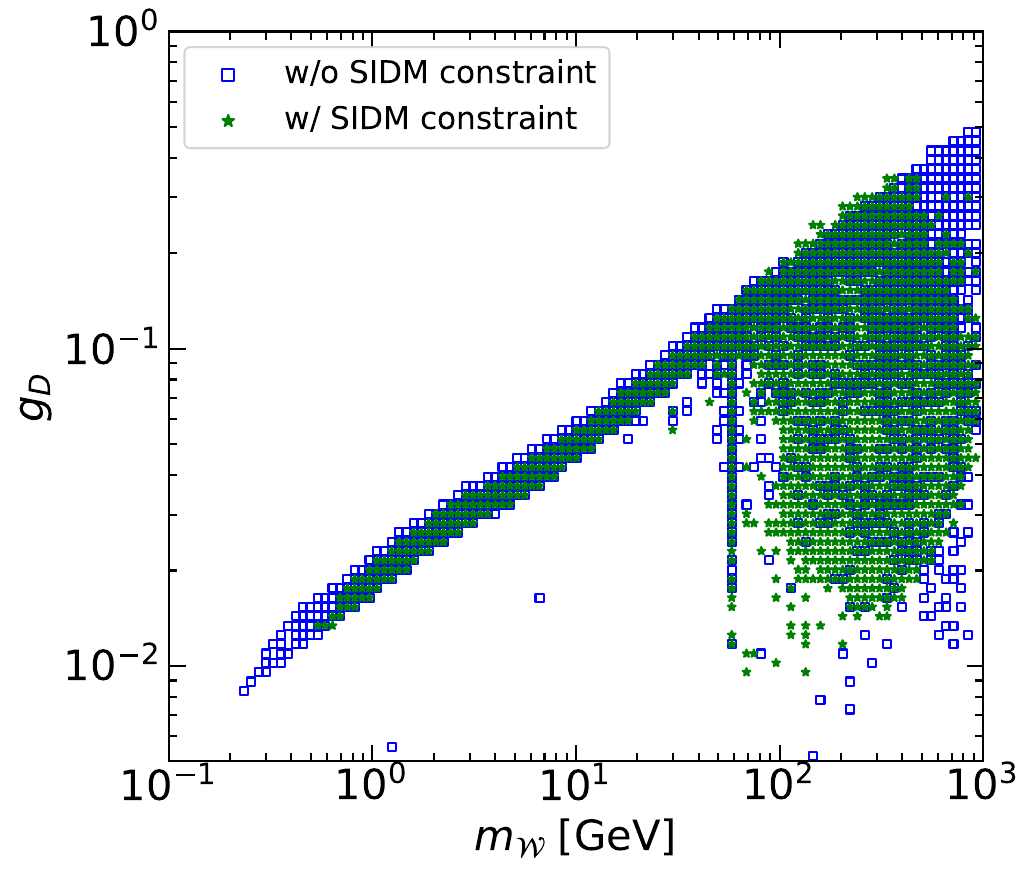}
    \includegraphics[width=0.49\textwidth]{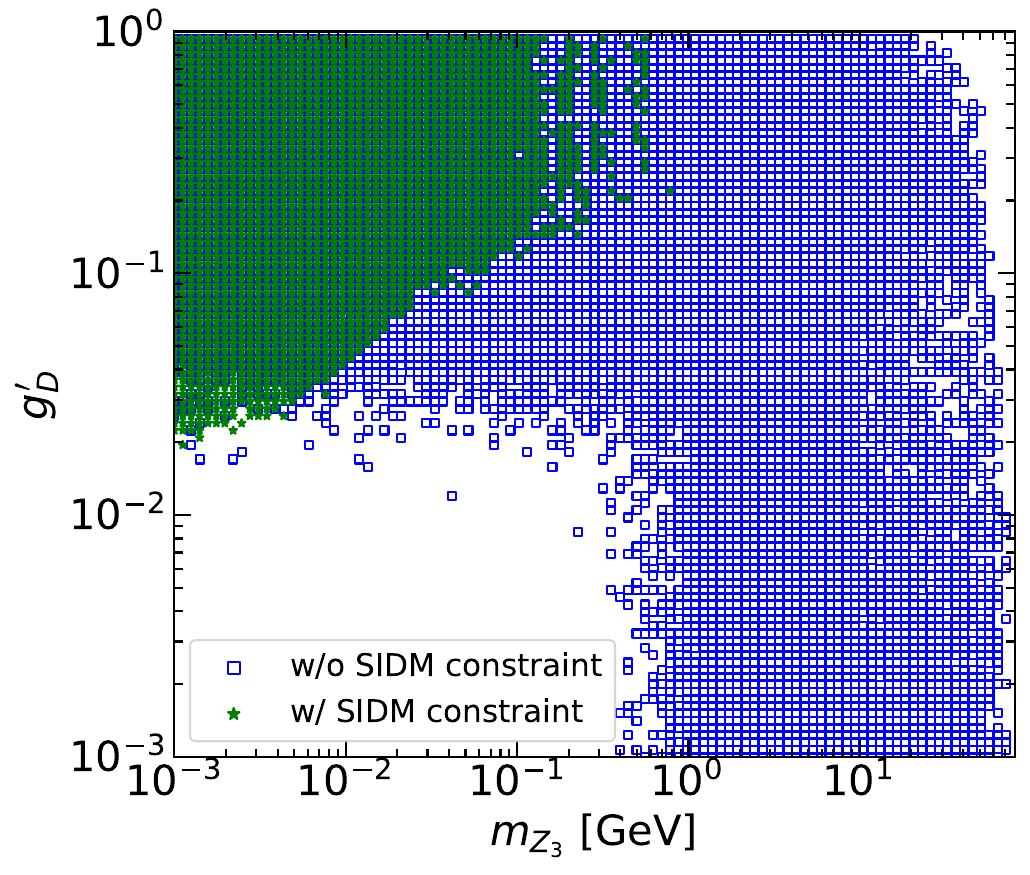}
    \caption{\label{fig:numerical_case2}The $2\sigma$ favored region projected on ($m_{\cal W}$, $g_D$) plane (left panel) and ($m_{Z_3}$, $g_D'$) plane (right panel) for the case $m_{Z_2} > m_{Z_3}$. The blue boxes indicate the favored region considering all model constraints except the self-interacting DM, while the green stars represent the inclusion of self-interacting DM.} 
\end{figure}

Here we present the numerical result for the case where $m_{Z_2} > m_{Z_3}$. 
For this case the SIDM cross section
dominated by the $Z_3$ mediation diagram. The favored region spanned on the ($m_{\cal W}$, $g_D$) and ($m_{Z_3}$, $g_D^{\prime}$) planes are shown in the left and right panels of Fig.~\ref{fig:numerical_case2}, respectively. In comparison with the results for the $m_{Z_2} < m_{Z_3}$ scenario depicted in Fig.~\ref{fig:numerical_res1}, it requires a stronger upper bound on $g_D$ and a stronger lower bound on $g_D^{\prime}$ within the SIDM favored region.

\end{document}